\documentclass[preprints,article,accept,moreauthors,pdftex]{Definitions/mdpi} 

\firstpage{1} 
\pubvolume{xx}
\issuenum{1}
\articlenumber{5}
\pubyear{2019}
\copyrightyear{2019}
\history{Received: date; Accepted: date; Published: date}

\usepackage{amsthm}
\usepackage{amsfonts}
\usepackage{amssymb}
\usepackage{graphicx}
\usepackage{physics}
\usepackage{siunitx}
\usepackage{color}
\usepackage{mathrsfs}
\usepackage{hhline}
\usepackage{mathtools}
\usepackage{multirow}

\newtheorem{subdefinition}{Definition}[definition]

\newcommand{\be}{\begin{equation}}
\newcommand{\ee}{\end{equation}}

\newcommand{\vars}[1]{\mathcal{#1}}
\newcommand{\bigepsilon}{\mathscr{E}}
\newcommand{\channel}{\mathscr{C}}
\newcommand{\user}{\mathsf{U}}
\newcommand{\eve}{\mathsf{E}}

\newcommand{\db}{\mathsf{D}}
\newcommand{\enc}{{\rm{enc}}}
\newcommand{\dec}{{\rm{dec}}}

\newcommand{\querya}{f_{\rm{query,1}}}
\newcommand{\queryb}{f_{\rm{query,2}}}

\newcommand{\answera}{f_{\rm{ans,1}}}
\newcommand{\answerb}{f_{\rm{ans,2}}}
\newcommand{\baranswera}{\bar{f}_{\rm{ans,1}}}
\newcommand{\baranswerb}{\bar{f}_{\rm{ans,2}}}
\newcommand{\decode}{f_{\rm{dec}}}
\newcommand{\advstate}{\sigma}
\newcommand{\btwoprot}{\mathscr{B}_2''}

\newcommand{\abort}{\perp}
\newcommand{\epscorr}{\varepsilon_{\textrm{cor}}}
\newcommand{\epssec}{\varepsilon_{\textrm{sec}}}
\newcommand{\rhoQKD}{\rho^{\textrm{real}}}
\newcommand{\rhoideal}{\rho^{\textrm{ideal}}}
\newcommand{\pass}{\textrm{pass}}
\newcommand{\fail}{\textrm{fail}}
\newcommand{\CDS}{\textrm{CDS}}
\newcommand{\corr}{\textrm{cor}}
\newcommand{\UP}{\textrm{UP}}
\newcommand{\DP}{\textrm{DP}}
\newcommand{\PS}{\textrm{PS}}
\graphicspath{{Images/}}

\Title{Provably-secure symmetric private information retrieval with quantum cryptography}

\Author{Wen Yu Kon $^{1,*}$\orcidA{} and Charles Ci Wen Lim $^{1,2}$\orcidB{}}
\AuthorNames{Wen Yu Kon and Charles Ci Wen Lim}

\address{
$^{1}$ \quad Department of Electrical \& Computer Engineering, National University of Singapore, Singapore\\
$^{2}$ \quad Centre for Quantum Technologies, National University of Singapore, Singapore
}

\corres{Correspondence: wkon01@u.nus.edu.sg}

\abstract{\emph{Private information retrieval} (PIR) is a database query protocol that provides user privacy, in that the user can learn a particular entry of the database of his interest but his query would be hidden from the data centre. \emph{Symmetric private information retrieval} (SPIR) takes PIR further by additionally offering database privacy, where the user cannot learn any additional entries of the database. Unconditionally secure SPIR solutions with multiple databases are known classically, but are unrealistic because they require long shared secret keys between the parties for secure communication and shared randomness in the protocol. Here, we propose using quantum key distribution (QKD) instead for a practical implementation, which can realise both the secure communication and shared randomness requirements. We prove that QKD maintains the security of the SPIR protocol and that it is also secure against any external eavesdropper. We also show how such a classical-quantum system could be implemented practically, using the example of a two-database SPIR protocol with keys generated by measurement device-independent QKD. Through key rate calculations, we show that such an implementation is feasible at the metropolitan level with current QKD technology.}


\begin{document}

\section{Introduction}

With the rising concern of personal data privacy, users of digital services may not want their preferences or selections to be revealed to service providers. This can be achieved with \emph{private information retrieval} (PIR), where users can access specific entries of a database held by the service provider at a data centre without revealing his or her entry selection~\cite{Chor1998}. This cryptographic technique has found application in areas such as anonymous communication~\cite{Mittal2011} and protecting user location privacy in location-based services~\cite{Khoshgozaran2008}.

However, in some occasions, the service provider or data centre may not want to reveal more information about the database than what is necessary, i.e., than what should have been given to the user. Such a setting is common in pay-per-access platforms such as iTunes and Google Play, or in more sensitive environments where the service provider has to secure the information of other database entries, like in the case for medical records retrieval and biometrics authentication~\cite{Bringer2007}. To provide for this additional security requirement (i.e., database privacy), one may employ \emph{symmetric private information retrieval} (SPIR), a sort of two-way secure retrieval scheme first introduced by Gertner \emph{et. al}.~\cite{Gertner2000}. 

In the literature, both PIR and SPIR have been extensively studied in the case where the user only communicates with one data centre. Here in the former, unconditional security (or information-theoretic security) can only be achieved by communicating the entire database from the data centre to the user. This implies that information-theoretic single database SPIR is not achievable~\cite{Chor1998}. To overcome this impasse, researchers have looked to weaker security frameworks, for instance, those based on computational security~\cite{Stern1998,Lipmaa2005,Naor2001,Chou2015}. 

On the quantum front, there is also a similar conclusion for single database SPIR~\cite{Lo1997}, i.e., it is not possible to achieve information-theoretic security even in the quantum setting. In light of these negative results, protocols for SPIR have largely evolved to cheat-sensitive protocols, also known as quantum private query~\cite{Giovannetti2008}. Examples of these protocols include those based on quantum oblivious key distribution~\cite{Jakobi2011,Rao2013,Zhang2013,Wei2016,Wei2020}, those based on sending states to a database oracle~\cite{Giovannetti2008Proof,Olejnik2011}, and those based on round-robin QKD protocol~\cite{Li2016}. In these protocols, the parties are averse to being caught cheating, so cheat-detection strategies allows one to construct protocols with more relaxed conditions as compared to those of SPIR~\cite{Gao2019}. However, parties can stand to gain information by cheating in these protocols and thus the protocols would not satisfy the original security requirements of SPIR proposed by Ref.~\cite{Gertner2000}. Other attempts at avoiding the no-go results include using special relativity~\cite{Kent2012,Garcia2016}.

One way to achieve information-theoretic security for SPIR is to communicate with multiple data centres, each of which holds a copy of the database. In fact, in their seminal work, Gertner \emph{et. al.} introduced a $k$-database classical SPIR protocol that is information-theoretically secure, with the assumption that the data centres cannot communicate (during and after the protocol), and how one can build these from $k$-database PIR protocols~\cite{Gertner2000}. Since then, researchers have studied multi-database SPIR in the context of compromised and byzantine data centres~\cite{Wang2017}. With multiple databases, the communication complexity of PIR and SPIR protocol can also be reduced to $O(n^{\frac{1}{2k-1}})$ based on Gertner's original protocol~\cite{Gertner2000}, and even further to $O(n^{10^{-7}})$ by Yekhanin~\cite{Yekhanin2008}, where $n$ is the number of entries in the database. There have also been several studies on the quantum version of multi-database SPIR. Kerenidis \emph{et. al.} focuses on how SPIR can be performed without shared randomness if the user is honest~\cite{Kerenidis2004}. Song et. al. proposed a quantum multi-database SPIR, but requires shared entanglement between the data centres and assumes secure classical and quantum channels~\cite{Song2019}.

The classical multi-database SPIR protocols proposed require secure channels, which are not achievable without some pre-shared secret keys between the parties in the protocol. In principle, the secret keys should be as long as the messages to be exchanged, but this would be costly and impractical for applications that work with large databases or require multiple uses of the SPIR protocol, e.g., medical records query where each doctor has to query for the files of multiple patients. 
In practice, the standard approach is to use public-key cryptography (e.g., using the Diffie–Hellman key distribution protocol~\cite{DH1976}) to expand the initial pre-shared secret key to a longer key. However, taking this approach could be risky, for it has been demonstrated that most known key distribution schemes based on public-key cryptography are insecure against quantum computing based attacks (an emerging technology). Evidently, this can be a huge problem for applications which require long-term security, like in the case of electronic health records which typically requires decades of information confidentiality. 

Quantum key distribution (QKD), a relatively mature technology with already multiple companies selling commercial QKD devices, offers a solid and promising solution to the above as it provides an information-theoretic method to expand pre-shared secret keys~\cite{BB1984,Gisin2002}. As such, the expanded keys can withstand the threats of quantum computing based attacks, and any other yet-to-be-discovered algorithmic advancements. 
Moreover, the expansion of keys need not be performed in real-time, i.e., expanded keys can be used for future SPIR runs. It is important to emphasise here that QKD cannot lead to a perfectly secure channel in practice, for it involves some statistical and entropy estimation procedures which carry overhead penalties in the security. Fortunately, these penalties can be made arbitrarily small with a proper security analysis, and subsequently the resulting secure channel can be made arbitrarily close to a perfect one. It is the goal of our work to incorporate these technical subtleties into the original security definition of SPIR so that we can add QKD as a supporting base layer. Here, we see the QKD layer as one that provides the necessary secret keys and secure channels (using one-time pad encryption) for SPIR. We note that Quantum Secure Direct Communication, which transmits messages directly using quantum states, could also serve as a secure communication channel~\cite{Deng2003,Zhu2017,Qi2019}.

\begin{figure}[H]
    \centering
    \includegraphics[width=0.7\textwidth]{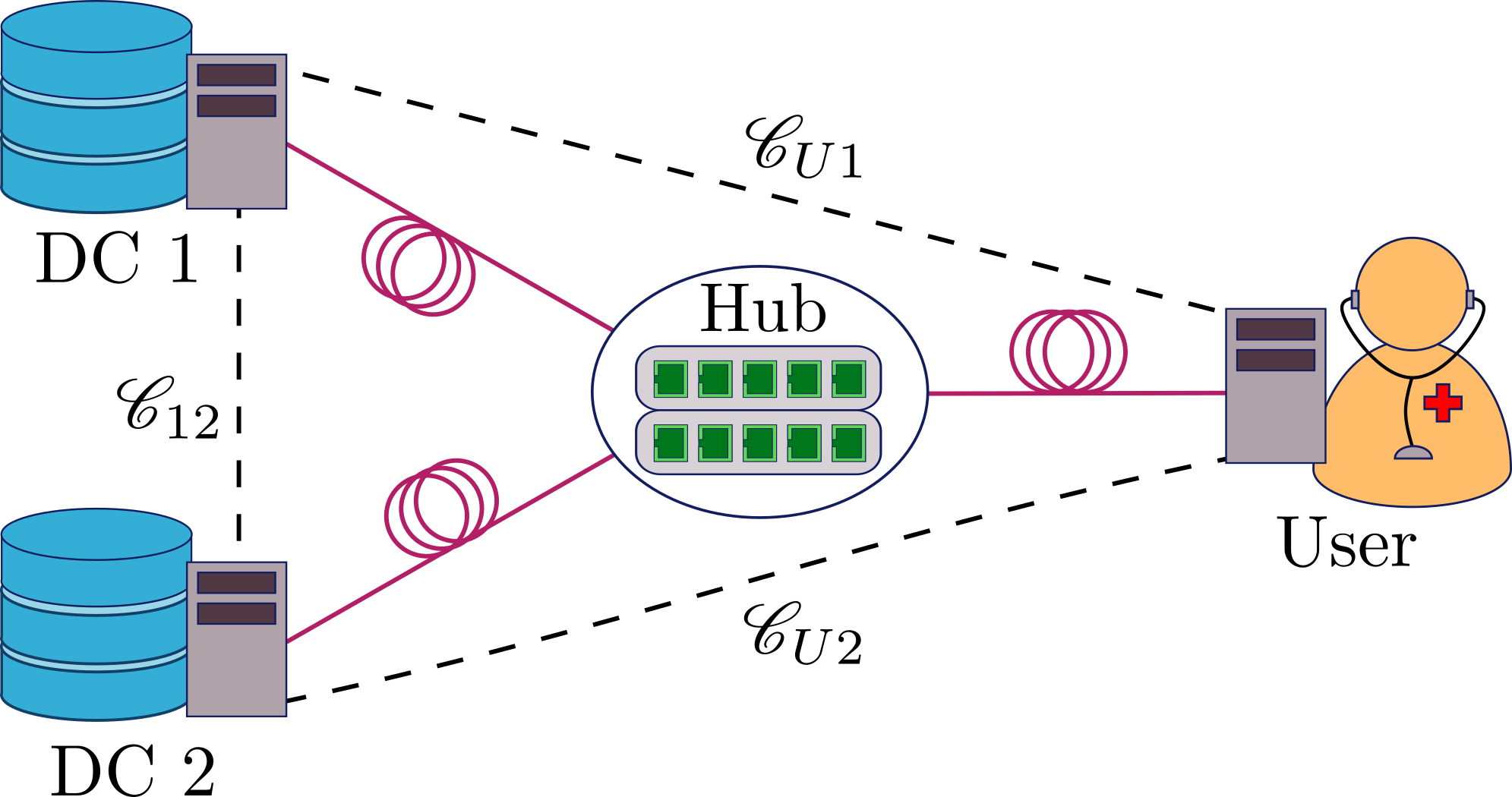}
    \caption{Schematic of a QKD network with star topology, which can supply QKD keys for the SPIR protocol. The central node (hub) connects to the user and two data centres with optical fibre (solid lines). Using the physical connection, any two parties in the protocol can establish a secure QKD link (dotted lines) via the central node.}
    \label{SetupFig}
\end{figure}

A widely studied QKD network configuration is the star topology, where multiple QKD users are connected to a (possibly untrusted) central node, and any two users can achieve secure communication by performing measurement device-independent (MDI) QKD~\cite{Lo2012,Liu2013,Yin2016,Tang2016} via the measurement device held by the central node. This choice of QKD network has the additional benefit that the number of quantum channels scales linearly in the number of users, which is an important consideration for practical deployments. To illustrate how SPIR can be implemented on this network, we turn to the example of accessing electronic health records on a database~\cite{EHR_Sec}. In this situation, we assume that the data centres (holding onto the health records) belongs to the health ministry, the user is a doctor in a government hospital wanting to query the medical records of a patient, and they are connected via a network service provider. As shown in Figure~\ref{SetupFig}, the network service provider holds the central node that connects to two data centres and the user in a star topology with optical fibre connections labelled by solid lines. Using MDI-QKD, any two parties can establish a secure QKD link via the central node, and these are labelled by dotted lines. The keys generated from these QKD links can then be used to establish secure communication for the classical SPIR protocol using one-time pad encryption. The doctors would thus be able to protect their patients' privacy when querying, and the health records of other patients held by the health ministry would remain private from both the querying doctor and the network service provider.

In this work, we describe how QKD can be used to relax the requirement of perfectly secure channels in classical multi-database SPIR, and examine the resources required for such a protocol. 
In Sec.~\ref{Prelim}, we introduce the mathematical notations required to describe the protocol and security analysis.
In Sec.~\ref{SPIR}, we introduce the basic elements of a generic SPIR protocol and the original SPIR security definition. 
In Sec.~\ref{QKD}, we introduce QKD channels and its security definitions, generalise the SPIR definition to a quantum one, and show how QKD can be incorporated into SPIR as the communication channel. 
In Sec.~\ref{SecProof}, we prove the security for a multi-database SPIR protocol with QKD channels based on the revised SPIR definitions. 
In Sec.~\ref{NumSim}, we introduce MDI-QKD and perform numerical analysis to determine the resources required for MDI-QKD to obtain the desired SPIR protocol.
 

\section{Preliminaries}
\label{Prelim}

\subsection{Quantum and Classical Systems}
The state of a generic quantum system living in Hilbert space $A$ is represented by a density operator $\rho_A$, a positive semi-definite matrix with trace one. Classical systems are modelled by quantum systems whose state is diagonal in a given orthonormal basis. For a random variable $Y$ that takes on values $y\in\vars{Y}$ with probability $P_Y(y)=\Pr[Y=y]$, the corresponding state of the classical random variable is

\begin{equation}
    \rho_Y=\sum_{y\in\vars{Y}}P_Y(y)\dyad{y},
\end{equation}
where $\{\ket{y}\}_{y\in\vars{Y}}$ forms an orthonormal basis. To keep the above notation compact for multiple variables, we will sometimes use $\Pi_{XYZ}(xyz)$ to represent the tensor product of classical states, i.e., $\dyad{x}\otimes\dyad{y}\otimes \dyad{z}$. 

A bipartite system on $YA$ is called \emph{classical-quantum} if its state admits the form

\begin{equation}
    \rho_{YA}=\sum_{y\in\vars{Y}}p_y\dyad{y}\otimes\rho_A^{y},
\end{equation}
where $\rho_A^{y}$ is the state of $A$ conditioned on the event $Y=y$. 

\subsection{Trace distance and distinguishability}

To measure the distinguishability of two quantum systems, we use the trace distance measure, which for any two states $\rho$ and $\sigma$, is defined as

\begin{equation}
    \Delta(\rho,\sigma)=\frac{1}{2}\norm{\rho-\sigma}_1,
\end{equation}
where $\norm{\rho-\sigma}_1$ is the trace norm of $\rho-\sigma$. Notice that the trace distance is bounded between 0 and 1, with identical states giving 0 and completely orthogonal states giving 1. With this, two systems are said to be $\varepsilon$-\emph{close} if their states, $\rho$ and $\sigma$, satisfy $\Delta(\rho,\sigma)\leq \epsilon$. The trace distance measure admits a few properties: (1) it satisfies triangle inequality, i.e., for any $\rho$, $\sigma$, and $\tau$, it satisfies $\Delta(\rho,\sigma)\leq\Delta(\rho,\tau)+\Delta(\tau,\sigma)$, (2) it is jointly convex in its inputs, i.e., $\Delta(\sum_i \lambda_i\rho_i,\sum_i \lambda_i\sigma_i)\leq\sum_i\lambda_i\Delta(\rho_i,\sigma_i)$, where $\lambda_i\geq 0$ and $\sum_i \lambda_i=1$, (3) it is non-increasing under completely positive and trace preserving (CPTP) maps $\bigepsilon$, i.e., $\Delta(\bigepsilon(\rho),\bigepsilon(\sigma))\leq\Delta(\rho,\sigma)$. For classical random variables $Y_1$ and $Y_2$ that takes on values $y\in\vars{Y}$ with probability distribution $P_{Y_1}$ and $P_{Y_2}$, the trace distance of their probability distributions reduces to the classical definition,

\begin{equation}
    \Delta(Y_1,Y_2)=\frac{1}{2}\sum_{y\in\vars{Y}}\abs{\Pr[Y_1=y]-\Pr[Y_2=y]}.
\end{equation}
If the random variables $Y_1$ and $Y_2$ correspond to the measurement outcome when performing a POVM measurement $\{\Gamma_y\}_{y\in\vars{Y}}$ on states $\rho$ and $\sigma$ respectively, the trace distance of the probability distribution of $Y_1$ and $Y_2$ would be upper bounded by the trace distance of the original quantum states~\cite{Nielsen2011}, i.e.

\begin{equation}
    \Delta(Y_1,Y_2)\leq\Delta(\rho,\sigma).
\end{equation}


\section{SPIR}
\label{SPIR}

\subsection{Generic one-round SPIR protocol}
\label{GenSPIRProtocolSec}

In this section, we introduce some additional notations and the essential elements of a generic SPIR protocol. A multi-database SPIR protocol has a user $\user$, who interacts with $k\geq 2$ data centres $\db_j$, $j\in\{1,\dots,k\}$, each having a copy of the database, represented by $W$ with $n$ entries. For simplicity, we focus on databases with single bit entries, i.e., $W=(W_1,W_2,\ldots,W_n)\in\{0,1\}^n$; our analysis can be easily extended to multi-bit entries. 

We also assume that all parties are equipped with a secure random number generator, which they may use for cryptography purposes. For our analysis, we denote the user's local randomness by $R$.

Here, we focus on one-round SPIR protocols, where there is only one round of query from the user to the data centres, and a single round of reply from the data centres to the user. In the case of multi-round SPIR protocols, there can be multiple successive rounds of queries and answers. A one-round SPIR protocol for two data centres can thus be defined by a pair of query functions, $\querya$ and $\queryb$, to generate the user queries for data centre 1 and data centre 2, respectively, answer functions $\answera$ and $\answerb$ for the data centres to generate their responses to the queries received, and the decoding function $\decode$ for the user to retrieve the desired database entry, $W_X$. These are functions of random variables and hence their outputs are random variables as well. 

A generic one-round two-database SPIR protocol typically performs the following steps (summarised in Table~\ref{GenSPIRTable}) for a given input $X=x$ and database $W=w$:
\begin{enumerate}[leftmargin=*,labelsep=4.9mm]
    \item \textbf{Establishing secure channels:} Using pre-established secret keys, perfectly secure channels are established between the user and data centres using one-time pad (OTP) encryption. We use $(K_1,K_2)$, $(K_3,K_4)$, $(K_5,K_6)$ to represent the secret key pair between data centre 1 and user, between data centre 2 and user, and between the data centres, respectively. For example, with this arrangement, the user holds $K_2$ and $K_4$ and data centre 1 holds $K_1$ and $K_5$. Secure channels connecting the user and data centres are denoted by $\channel_{U1}$ and $\channel_{U2}$, respectively. Note that the data centres are not allowed to communicate and hence we do not need to define any channel for them. To allow for two-way secure communication with a single secret key, we split $K=(K^{\enc},K^{\dec})$ into two halves, namely $K^{\enc}$ (for encryption) and $K^{\dec}$ (for decryption).
    
    \item \textbf{Query:}~The user generates queries for data centres 1 and 2, with $Q_1=\querya(x,R)$ and $Q_2=\queryb(x,R)$, respectively, and sends them to the data centres using the secure channels $\channel_{U1}$ and $\channel_{U2}$.
    
    \item \textbf{Answer:}~Upon receiving the query $\tilde{Q}_1$ (which could be different from $Q_1$), $\db_1$ (resp. $\db_2$) determines a reply $A_1=\answera(\tilde{Q}_1,w,K_5)$ (resp. $A_2=\answerb(\tilde{Q}_2,w,K_6)$ and sends it to the user via the secure channels.
    
    \item \textbf{Retrieval:}~The user retrieves the desired database entry value using $\hat{w}_x=\decode(\tilde{A}_1,\tilde{A}_2,Q_1,Q_2,x,R)$.
    
\end{enumerate}

SPIR is designed to resolve situations where the user or data centres deviate from their expected (honest) behaviour. For instance, a dishonest user could communicate bad queries in an attempt to learn additional entries in $w$, and dishonest data centres could provide replies other than the expected answer $A_j$ to learn about $x$. That is, a dishonest user can replace $Q_j$ in step 2 of the protocol by an adversarial query $\bar{Q}_j$, and dishonest data centres can provide adversarial answers $\bar{A}_j$ in step 3 of the protocol. 

Therefore, a secure SPIR protocol has to address both forms of attacks. At the heart of multi-database SPIR is the availability of pre-shared secret keys, which are pre-distributed between the users and the data centres. With these pairwise secret keys, the user can securely send his/her queries, $Q_1$ and $Q_2$, to the respective data centres, such that neither of the data centres can get both queries at the same time. Then, by also not allowing the data centres to communicate, one can enforce that neither of them can guess correctly $x$. Crucially, the use of secure channels also guarantees that no eavesdropper can get both $Q_1$ and $Q_2$ and hence $x$. These arguments collectively imply user privacy.

In the answer phase, it is important that the data centres do not reveal more than what is supposed to be given to the user. To achieve this, Gertner \emph{et. al.}~\cite{Gertner2000} introduced the task of \emph{conditional disclosure of secrets} (CDS). This is broadly described by a three-party task, where Alice and Bob, each with inputs $y$ and $z$, are supposed to reveal a common secret $s$ to Charlie, if and only if $y$ and $z$ satisfy a certain public predicate $f(y,z)$. Indeed, using this task, one could then draw immediate connections and see that $Q_1$ and $Q_2$ correspond to $y$ and $z$, respectively, and the common secret is the desired database entry $w_x$. Hence, for CDS to work, some private shared randomness between the data centres is necessary and this is exactly given by the secret key pair $(K_5,K_6)$. These arguments thus imply that the user cannot get the correct secret if the queries are not the expected ones, which in turn provides the required database privacy.

\begin{table}[H]
    \caption{\label{GenSPIRTable}Generic one-round two-database SPIR protocol.}
    \centering
    \tablesize{\scriptsize}
    \begin{tabular}{lccccc}
        \toprule
        \textbf{Step} & \textbf{$\db_1$} & & \textbf{$\user$} & & \textbf{$\db_2$} \\[5pt]
        \midrule
        Input: & $w$ & & $R$, $x$ & & $w$\\
         & & & & \\
        Key pair ($\db_1\leftrightarrow\db_2$): & $K_5$ & \multicolumn{3}{c}{$\xleftrightarrow{\mathmakebox[25em]{}}$} & $K_6$\\[5pt]
        Key pair ($\user\leftrightarrow\db_1$): & $K_1$ & $\xleftrightarrow{\mathmakebox[3em]{}}$ & $K_2$ & & \\[5pt]
        Key pair ($\user\leftrightarrow\db_2$): & & & $K_4$ & $\xleftrightarrow{\mathmakebox[3em]{}}$ & $K_3$\\[5pt]
        Query: & & & $Q_1=\querya(x,R)$,\quad $Q_2=\queryb(x,R)$ & & \\[5pt]
        OTP ($\user\rightarrow\db_1$): & $\tilde{Q}_1=C_{Q_1}\oplus K_1^\dec$ & $\xleftarrow{\mathmakebox[3em]{\channel_{U1}}}$ & $C_{Q_1}=Q_1\oplus K_2^{\enc}$ & & \\[5pt]
        OTP ($\user\rightarrow\db_2$): & & & $C_{Q_2}=Q_2\oplus K_4^{\enc}$ & $\xrightarrow{\mathmakebox[3em]{\channel_{U2}}}$ & $\tilde{Q}_2=C_{Q_2}\oplus K_3^{\dec}$\\[5pt]
        Answer: &  $A_1=\answera(\tilde{Q}_1,w,K_5)$ & & & & $A_2=\answerb(\tilde{Q}_2,w,K_6)$ \\[5pt]
        OTP ($\db_1\rightarrow\user$): & $C_{A_1}=A_1\oplus K_1^{\enc}$ & $\xrightarrow{\mathmakebox[3em]{\channel_{U1}}}$ & $\tilde{A}_1=C_{A_1}\oplus K_2^{\dec}$ & & \\[5pt]
        OTP ($\db_2\rightarrow\user$): & & & $\tilde{A}_2=C_{A_2}\oplus K_4^{\dec}$ & $\xleftarrow{\mathmakebox[3em]{\channel_{U2}}}$ & $C_{A_2}=A_2\oplus K_3^{\enc}$ \\[5pt]
        Decoding: & & & $\hat{w}_x=\decode(\tilde{A}_1,\tilde{A}_2,Q_1,Q_2,x,R)$ & & \\
        \bottomrule
    \end{tabular}
\end{table}

\subsection{Original SPIR security definition}

At this point, it is useful to recap the original security definitions introduced by Gertner \emph{et. al.}~\cite{Gertner2000}. A SPIR protocol is said to be secure if it satisfies the \emph{correctness}, \emph{user privacy}, and \emph{database privacy} conditions. Since the setting here is purely classical, we assume that the output views are simply represented by random variables. More concretely, the view of the user is modelled by random variable $V_{\user}^w$, and the view of the data centre $j$ is modelled by $V_{\db_j}^x$, for $j=1,2$, where the dependence of $V_{\user}$ (resp. $V_{\db_j}$) on $w$ (resp. $x$) is explicitly labelled. Evidently, $V_{\user}$ also contains query information, $Q_1$ and $Q_2$, and communicated answers $\tilde{A}_1$ and $\tilde{A}_2$, while $V_{\db_j}$ contains $\tilde{Q}_j$ and $A_j$, for example. 
\setcounter{definition}{1}
\begin{subdefinition}[Correctness]
\label{OriginalCorr}
    When all the parties in the protocol are honest, then for any database query $x$ and database $w$, the protocol outputs $\hat{w}_x=w_x$.
\end{subdefinition}
\begin{subdefinition}[User Privacy]
\label{OriginalUserPriv}
    When the user is honest, then for any $w$ and $k_5$ (or $k_6$), and for all $x$ and $x'$, each data centre's view satisfies $\Delta(V_{\db_j}^x,V_{\db_j}^{x'})=0$.
\end{subdefinition}
\begin{subdefinition}[Database Privacy]
\label{OriginalDBPriv}
    When the data centres are honest, then for any $x$ and $r$, there exist an $x'$ such that for all $w$ and $w'$ with $w_{x'}=w_{x'}'$, the view of the user satisfies $\Delta(V_{\user}^w,V_{\user}^{w'})=0$.
\end{subdefinition}

The definition of correctness ensures that the protocol yields the desired result $w_x$ for the user. For user privacy, the trace distance measure is used as a distance metric for measuring the distinguishability of the views. To see this, suppose a hypothetical experiment where the data centre is randomly given two views, $V_{\db_j}^x$ and $V_{\db_j}^{x'}$, and has to determine which of the views is given to him. His maximum probability of guessing correctly the identity is directly linked to the trace distance, i.e., $1/2+\Delta(V_{\db_j}^x,V_{\db_j}^{x'})/2$. From this expression, it is then clear that the trace distance quantifies the advantage the data centre has in distinguishing between $V_{\db_j}^x$ and $V_{\db_j}^{x'}$. Hence, having zero advantage in distinguishing between a system with $x$ and one with $x'$ indicates that the data centre can gain no information about $X$. For database privacy, a dishonest user can input any $x$, since the adversarial queries $\bar{Q}_1$ and $\bar{Q}_2$ may not depend on this particular choice of $x$. 
For instance, a dishonest user can use his local randomness $R$ to choose queries $\bar{Q}_1$ and $\bar{Q}_2$ that corresponds to queries for different $x$. For each $r$ (i.e. each possible choice of queries), the information to which the user truly intends to learn would be implicitly carried by $\bar{Q}_1$ and $\bar{Q}_2$.
Therefore, the existence of an $x'$ such that the user cannot distinguish between $w$ and $w'$ satisfying $w_{x'}=w'_{x'}$ for each $r$ means that the user is unable to obtain any information beyond a single entry of the database, $w_{x'}$, for whichever queries that is randomly selected for that run. 


\section{SPIR with QKD}
\label{QKD}
\subsection{QKD channel}
As mentioned above, our goal is to replace the perfectly secure communication channels assumed in multi-database SPIR with QKD channels. Before going into more details, it is useful to first recap the essential features of QKD and its security definitions. 

The goal of QKD is to generate a pair of secure keys which are identical, uniform and secret from any eavesdropper. In this setting, we assume that the underlying QKD devices are honest and they each have a trusted local source of randomness. Below, we use random variable $S$ instead of $K$ to represent QKD keys.  

A generic QKD between party $A$ and party $B$ can either succeed in producing a pair of keys, $S_A,S_B\in\vars{S}$ (with probability $1-p_{\abort}$), or abort and output an abort flag, $S_A=S_B=\,\perp$ (with probability $p_{\abort}$). The average output state of a QKD protocol is hence given by

\be
    \rhoQKD_{S_AS_BE}=p_{\abort}\Pi_{S_AS_B}(\perp\perp)\otimes\sigma_E^{\perp} + \sum_{s,s'\in \vars{S}} P_{S_AS_B}(s,s') \Pi_{S_AS_B}(ss')\otimes \sigma_E^{s,s'},
\ee
where $p_{\abort}=P_{S_AS_B}(\perp,\perp)$ is the abort probability and $\advstate_E^{s,s'}$ is the quantum state conditioned on the outcome $(s,s')$ held by an eavesdropper at the end of the protocol. For brevity, we shall use $\perp$ to label a normalised state that is conditioned on protocol aborting, and $\top$ to label a normalised state that is conditioned on the protocol not aborting. For instance, in the above equation, the first term corresponds to $p_{\perp}\rho_{S_AS_BE}^{\rm{real},\perp}$, and the second term corresponds to $(1-p_{\perp})\rho_{S_AS_BE}^{\rm{real},\top}$.

\subsection{QKD security definition}

Keys generated from QKD may not be perfectly uniform and secret from the eavesdropper, but one can ensure that the keys are asymptotically close (in trace distance) to an ideal key by choosing the right security parameter. This security parameter is defined by the distinguishability of QKD keys from an ideal key.
The ideal key described here is related, but slightly different from the secret key utilised for a secure classical channel. Since QKD channels can abort, the ideal key used for comparison has probability $p_{\abort}$ of returning an abort flag, whereas the process of sharing secret keys for secure channels are typically assumed not to fail. This introduces a loss in the \emph{robustness} of the channel (i.e. it can sometime fail), but does not compromise channel security since protocol aborting does not provide Eve with any information on the message.
The ideal output state of a QKD is given as

\be
    \rho^{\rm{ideal}}_{S^{AB}E}= p_{\abort}\Pi_{S_AS_B}(\perp\perp)\otimes\sigma_E^{\perp} + \frac{1}{\abs{\vars{S}}} \sum_{s,s'\in \vars{S}:s=s'} \Pi_{S_AS_B}(ss')
        \otimes \advstate_E,
\ee
where $\advstate_E=\sum_{s{''},s{'''}\in \vars{S}} P_{S_AS_B}(s'',s{'''})\sigma_E^{s'',s'''}$ is the marginal state of Eve. 

Following Ref.~\cite{Portmann2014}, a QKD protocol is said to be $\varepsilon$-secure if the actual QKD and ideal output states satisfy

\begin{equation}
    \Delta(\rhoQKD_{S_AS_BE},\rho^{\rm{ideal}}_{S_AS_BE})\leq\varepsilon.
\end{equation}
The security of QKD can, in fact, be seen as the sum of two security criteria, namely \emph{correctness} and \emph{secrecy}. More specifically, it can be shown that, 

\be
 \Delta(\rhoQKD_{S_AS_BE},\rho^{\rm{ideal}}_{S_AS_BE}) \leq \Pr[S_A\neq S_B] + (1-p_\abort)\Delta(\rhoQKD_{S_AE},\rhoideal_{S_AE}),
\ee
where the terms on the R.H.S. are the correctness and secrecy conditions, respectively, and they satisfy

\begin{equation}
\begin{split}
    \Pr[S_A\neq S_B] \leq \epscorr,\\ (1-p_\abort)\Delta(\rhoQKD_{S_AE},\rhoideal_{S_AE}) \leq \epssec.
\end{split}
\end{equation}
These criteria imply that $\varepsilon = \epscorr+\epssec$. 

The correctness criterion, in practice, is typically enforced by using hashing, which guarantees that the two keys are identical except with some small error probability, $\epscorr/(1-p_\abort)$. That is, given the protocol does not abort, the maximum probability that the generated keys are different is given by $(1-p_{\abort})\Pr[S_A\neq S_B|\pass]\leq\epscorr$. The secrecy criterion looks at how distinguishable the output state of either $S_A$ or $S_B$ is from the ideal output, after passing through the privacy amplification step using a quantum-proof randomness extractor. For more details of these criteria, we refer the interested reader to Ref.~\cite{Portmann2014}. In the following, for simplicity, we assume that all QKD channels use the same security parameters, i.e., $\epscorr$ and $\epssec$, for these can be enforced in practice with the right error verification and privacy amplification schemes. The robustness probability is however harder to enforce as it depends on the quantum channel behaviour which can be different between channels. To that end, we will write $p_{\abort,U1}$, $p_{\abort,U2}$, and $p_{\abort,12}$ to represent the abort probabilities for QKD pairings $(\user,\db_1)$, $(\user,\db_2)$, and $(\db_1,\db_2)$, respectively. 

\subsection{SPIR with QKD security definition}
In order to analyse SPIR protocols that utilise QKD keys, it is necessary to generalise the original SPIR security definition. These changes will have to accommodate aspects of a QKD channel that are not normally present in a perfectly secure channel. More specifically, we need to consider the possibility that the QKD protocol can abort, and that it has a non-zero probability of outputting an imperfect secret key pair.

In the original SPIR setting, a two-party protocol between the data centres and user is considered. Here, no external eavesdropper is included, for secure channels are used and hence no external party can obtain any information from the communication. However, in the case of practical QKD systems, there is a small possibility that the eavesdropper could learn something about the secret keys. To allow for such bad events, we look at SPIR as a three-party protocol with an eavesdropper called Eve, and introduce a fourth condition which we term as \emph{protocol secrecy}. Similar to the other security conditions, the protocol secrecy condition requires that the view of any eavesdropper $E$ be independent of both $X$ and $W$, assuming both the user and data centres are honest. In the following, we first highlight four considerations when extending the original SPIR security definition to one that appropriately captures all possible bad events that may be caused by imperfect QKD keys.

Firstly, in analysing user privacy (resp. database privacy), the possibility of getting imperfect secret keys provides a new avenue for data centres (resp. the user) to gain more information on $X$ (resp. $W$). For instance, when the key pair $(S_3,S_4)$ is insecure, data centre 1 can gain information on $Q_2$ and $A_2$, which can be utilised to determine $x$. To suitably address these threats, we treat such situations as a collusion between the data centre and Eve (whose view contains the ciphertext). In other words, in analysing user privacy (resp. database privacy), we always assume that the dishonest party is colluding with the external eavesdropper, Eve.

Secondly, a feature of the current security definition of QKD is that the security error (the probability that the generated secret keys are imperfect/insecure) can be made arbitrarily small in the limit of infinitely long keys. To allow for this feature as well in the extended setting, we introduce compatible definitions by adding security parameters to each of the condition, all of which should be possible to make asymptotically small. For instance, the security parameter for correctness, $\eta_{\corr}$, would bound the probability of error in recovering $w_x$, the security parameters for user privacy, database privacy and protocol secrecy, $\eta_{\UP}$, $\eta_{\DP}$ and $\eta_{PS}$, would bound the difference between the two views given in the condition.

Thirdly, the possibility of having a mismatch of QKD keys for various communication channels would lead to inaccuracies when the classical SPIR definition is used. For user privacy, the classical definition requires the data centre's view to be independent of $X$ for any $k_5$, the shared random string between the databases. The definition also requires the same to be true for any $k_6$, but this need not be included since $K_5=K_6$ is assumed. Since QKD keys could be mismatched, $S_5\neq S_6$, $S_6$ has to be explicitly included in the adjusted definition. A similar problem is present for database privacy. The classical definition fixes $x$ and $r$, thereby fixing the adversarial queries $\bar{q}_1$ and $\bar{q}_2$ while analysing the user's view. This allows one to address any probabilistic strategy a dishonest user can perform by analysing each possible pair of query $\bar{q}_1$ and $\bar{q}_2$ that the user includes in his probabilistic strategy. If the user is unable to obtain more than $w_{x'}$ for some $x'$ for each pair of query, his probabilistic strategy would not yield more than a single entry of the database. Using QKD keys ($S_1^{dec}$,$S_2^{enc}$,$S_3^{dec}$,$S_4^{enc}$) can result in the queries $\tilde{\bar{Q}}_1$ and $\tilde{\bar{Q}}_2$ arriving at the databases being probabilistic, since there is a small probability that the keys do not match. For instance, $\bar{Q}_1$ and $\bar{Q}_2$ can be queries for $w_1$, but there is a small probability that the QKD keys are mismatched such that $\tilde{\bar{Q}}_1$ and $\tilde{\bar{Q}}_2$ queries for $w_2$, which means that there would not be an $x'$ for which the user's view is identical for any $w$ and $w'$ with $w_{x'}=w'_{x'}$. However, for each fixed set of QKD keys ($s_1^{dec}$,$s_2^{enc}$,$s_3^{dec}$,$s_4^{enc}$), the queries do indeed reveal at most a single $w_{x'}$ to the user. Therefore, the definition has to be adjusted to analyse the user's view with fixed keys ($s_1^{dec}$,$s_2^{enc}$,$s_3^{dec}$,$s_4^{enc}$).

Lastly, unlike secure communication channels, QKD protocols can fail due to reasons like high channel noise or failure to have matching hash values in the error verification step. In fact, even in the classical case, it is not inconceivable that an external party can perform denial of service attack on the channel, e.g., by physically cutting the optical channel. In such a situation, $w_x$ cannot be recovered and the correctness condition will not be met. To accommodate for such bad events, we modify the definition to condition out failure events (i.e. only consider `pass' cases), which has probability 

\be
    p_{fail}=1-(1-p_{\perp,U1})(1-p_{\perp,U2})(1-p_{\perp,12}).
\ee
This conditioning can be performed in practice since an abort flag, $\perp$, is sent in the case of protocol failure.
This is different from having an error in the decoded bit $\hat{w}_x$, which would be undetectable. Typically, once a QKD protocol aborts, the users will run the protocol again. However, for simplicity, we do not include this consideration in our analysis. Nevertheless, we remark that one should make $p_{fail}$ as small as possible in practice.

The extended security definitions are as follow:
\setcounter{definition}{2}
\setcounter{subdefinition}{0}
\begin{subdefinition}[$\eta_{\corr}$-correctness]
   Assuming the user and the data centres are honest, then for any $x$ and $w$, the protocol must fulfil $(1-p_{\fail})\Pr[\hat{w}_x\neq w_x|\pass]\leq \eta_{\corr}$.
\end{subdefinition}
\begin{subdefinition}[$\eta_{\UP}$-user privacy]
    \label{UserPrivDefn}
    Assuming the user is honest, then for any $w$ and shared keys between the databases ($s_5$,$s_6$), the total view of each data centre and the eavesdropper (Eve) has to fulfil $\Delta(\rho_{\db_j\eve}^x,\rho_{\db_j\eve}^{x'})\leq\eta_{\UP}$ for all $x$ and $x'$.
\end{subdefinition}
\begin{subdefinition}[$\eta_{\DP}$-database privacy]
    \label{DBPrivDefn}
    Assuming the data centres are honest, then for any $x$, $r$ and keys $(s_1^{dec},s_2^{enc},s_3^{dec},s_4^{enc})$, there exist an $x'$ such that for all $w$ and $w'$ with $w_{x'}=w'_{x'}$, the total view of the user and eavesdropper (Eve) has to fulfil $\Delta(\rho_{\user\eve}^w,\rho_{\user\eve}^{w'})\leq\eta_{\DP}$.
\end{subdefinition}
\begin{subdefinition}[$\eta_{\PS}$-protocol secrecy]
    Assuming the user and the data centres are honest, then for all $(x,w)$ and $(x',w')$, the view of the eavesdropper (Eve) has to fulfil $\Delta(\rho_{\eve}^{x,w},\rho_{\eve}^{x',w'})\leq\eta_{\PS}$.
\end{subdefinition}

We call any SPIR protocol that satisfies the above four conditions as ($\eta_{\corr}$,$\eta_{\UP}$,$\eta_{\DP}$,$\eta_{\PS}$)-secure. Note that the original SPIR definition can be recovered by taking (0,0,0,0)-security and assuming that there is no protocol failure $p_{\fail}=0$, that the shared random key between the databases are correct ($S_5=S_6$), and the user queries are communicated without errors ($S_1^{dec}=S_2^{enc}$ and $S_3^{dec}=S_4^{enc}$). More concretely, definition~\ref{OriginalCorr} is obtained since $\eta_{\corr}=0$ and $p_{\fail}=0$ implies $\Pr[\hat{w}_x\neq w_x]=0$, definitions~\ref{OriginalUserPriv} and \ref{OriginalDBPriv} are obtained by noting that the trace distance measure is contractive under partial trace operations.

\subsection{Quantum view modelling}

In Ref.~\cite{Gertner2000}, the authors proved that there exist a family of (0,0,0,0)-secure SPIR protocols assuming secure classical channels. However, establishing these secure channels require that the user and data centres have pre-shared keys that are at least as long as the messages to be sent. Pre-shared keys between the data centres are also required to perform CDS. This would be impractical for large databases or situations that require multiple uses of the SPIR protocol. Therefore, we can capitalise on QKD, which is a key expansion protocol. Starting with a small shared key between two parties, QKD can generate a much longer secret key for use. Hence, we establish QKD links between the parties to generate keys for both communication (between the user and data centres) and as shared randomness (between the data centres).

To analyse the security of the SPIR protocol with QKD, we need to first examine the view of various parties in the quantum setting. The protocol follows the generic one-round SPIR protocol described in Sec.~\ref{GenSPIRProtocolSec}, except that the keys used in key pairing steps are given by QKD keys instead. More specifically, we replace $(K_1,K_2)$, $(K_3,K_4)$, and $(K_5,K_6)$ by QKD generated keys $(S_1,S_2)$, $(S_3,S_4)$, and $(S_5,S_6)$, respectively. We also take that each set of QKD keys shared between two parties is generated by a single round of QKD. If any of the three QKD protocols aborts, i.e., if any of $(S_1,S_2)$, $(S_3,S_4)$ or $(S_5,S_6)$ returns $\perp$ after the first step of establishing secure channels, then the SPIR protocol will abort. For simplicity, we take that all random variables that are generated in the latter steps, including queries, answers and ciphertext, are set to $\perp$. The overall protocol is summarised in Table~\ref{QKDSPIRProtocolTable}.

By expressing the inputs as quantum states and steps in the protocol as maps, we can obtain the final state for all four parties, and determine each of their view by performing a partial trace. Here, we introduce four relevant views that are used in the SPIR security definition. The total view of the user and Eve (used in database privacy) is

\begin{equation} \label{usereveview}
\rho_{\user\eve}^w= \rho_{XRQ_1Q_2\tilde{A}_1\tilde{A}_2S_2S_4C_{Q_1}C_{Q_2}C_{A_1}C_{A_2}E}^w,
\end{equation}
the total view of Eve and data centre 1, and that of Eve and data centre 2 (used in user privacy) are

\begin{eqnarray}
\label{dc1eveview}
\rho_{\db_1\eve}^x&=&\rho_{W\tilde{Q}_1A_1S_1S_5C_{Q_1}C_{Q_2}C_{A_1}C_{A_2}E}^x,\\\label{dc2eveview}
\rho_{\db_2\eve}^x&=&\rho_{W\tilde{Q}_2A_2S_3S_6C_{Q_1}C_{Q_2}C_{A_1}C_{A_2}E}^x,
\end{eqnarray}
respectively, and the view of Eve (used in protocol secrecy) is

\begin{equation}
\label{eveview}
\rho_{\eve}^{x,w}=\sigma_{C_{Q_1}C_{Q_2}C_{A_1}C_{A_2}E}^{x,w}.
\end{equation}
Here, we note that $E$ is the side-information of Eve gathered up the OTP steps. As such, $E$ contains all of the quantum information exchanged over the QKD channels and all of the classical information exchanged due to error correction, verification, and privacy amplification.

\begin{table}[H] 
    \caption{\label{QKDSPIRProtocolTable}Generic one-Round two-database SPIR protocol with QKD}
    \tablesize{\scriptsize}
    \begin{tabular}{lcccccc}
        \toprule
        \textbf{Step} & \textbf{$\db_1$} & & \textbf{$\user$} & & \textbf{$\db_2$} & \textbf{$\eve$} \\[5pt]
        \midrule
        Input: & $w$ & & $R$, $x$ & & $w$ & \\[5pt]
         & & & & & \\
        QKD ($\db_1\leftrightarrow\db_2$): & $S_5$ & \multicolumn{3}{c}{$\xleftrightarrow{\mathmakebox[21em]{\rhoQKD_{S_5S_6E}}}$} & $S_6$ & $\sigma_{\eve}^{S_5S_6}$\\[5pt]
        QKD ($\user\leftrightarrow\db_1$): & $S_1$ & $\xleftrightarrow{\mathmakebox[3em]{\rhoQKD_{S_1S_2E}}}$ & $S_2$ & & &  $\sigma_E^{S_1S_2}$\\[5pt]
        QKD ($\user\leftrightarrow\db_2$): & & &  $S_4$ & $\xleftrightarrow{\mathmakebox[3em]{\rhoQKD_{S_3S_4}}}$ & $S_3$ & $\sigma_E^{S_3S_4}$\\[5pt]
        \multirow{2}{*}{Query:} & & & $Q_1=\querya(x,R)$ & & & \\[5pt]
        & & & $Q_2=\queryb(x,R)$ & & & \\[5pt]
        OTP ($\user\rightarrow\db_1$): & $\tilde{Q}_1=C_{Q_1}\oplus S_1^\dec$ & $\xleftarrow{\mathmakebox[3em]{\channel_{U1}}}$ & $C_{Q_1}=Q_1\oplus S_2^\enc$ & & & $C_{Q_1}$\\[5pt]
        OTP ($\user\rightarrow\db_2$): & & & $C_{Q_2}=Q_2\oplus S_4^\enc$ & $\xrightarrow{\mathmakebox[3em]{\channel_{U2}}}$ & $\tilde{Q}_2=C_{Q_2}\oplus S_3^\dec$ & $C_{Q_2}$\\[5pt]
        Answer: & $A_1=\answera(\tilde{Q}_1,w,S_5)$ & & & & $A_2=\answerb(\tilde{Q}_2,w,S_6)$ & \\[5pt]
        OTP ($\db_1\rightarrow\user$): & $C_{A_1}=A_1\oplus S_1^\enc$ & $\xrightarrow{\mathmakebox[3em]{\channel_{U1}}}$ & $\tilde{A}_1=C_{A_1}\oplus S_2^\dec$ & & & $C_{A_1}$\\[5pt]
        OTP ($\db_2\rightarrow\user$): & & & $\tilde{A}_2=C_{A_2}\oplus S_4^\dec$ & $\xleftarrow{\mathmakebox[3em]{\channel_{U2}}}$ & $C_{A_2}=A_2\oplus S_3^\enc$ & $C_{A_2}$\\[5pt]
        Decoding: & & & $\hat{w}_x=\decode(\tilde{A}_1,\tilde{A}_2,Q_1,Q_2,x,R)$ & & & \\[5pt]
        \bottomrule
    \end{tabular}
\end{table}


\section{Security analysis}
\label{SecProof}

Here, we show that the security parameters of the associated QKD protocols can be used to bound the generalised SPIR security parameters defined above.
\begin{Theorem}
    A two-database one-round $(0,0,0,0)$-secure SPIR protocol that uses $\varepsilon$-secure QKD keys in place of ideal keys, where $\varepsilon=\epscorr+\epssec$, is $\left(3\epscorr,2\varepsilon,2\varepsilon,4\varepsilon\right)$-secure.
\end{Theorem}

\emph{Proof sketch.---} For the correctness condition, if the all of the QKD key pairs are correct and conditioned on not aborting, then the 0-correctness of the SPIR protocol guarantees that the decoding will be correct. Moreover, since there may be key pair events other than the correct ones that can yield $\hat{w}_x=w_x$, we have that

\be
\label{CorrEqn}
    \Pr[\hat{w}_x=w_x|\pass]\geq \Pr\left[\{S_1=S_2\}\right.\cap\{S_3=S_4\}
    \left.\cap\{S_5=S_6\}|\pass\right],
\ee 
where the conditioning is that all of the QKD protocols do not abort. Then, by using the union bound, it is straightforward to show that the probability of error is upper bounded by the sum of the probability of each QKD key being wrong, and thus

\begin{equation}
    (1-p_{\fail})\Pr[\hat{w}_x\neq w_x|\pass]\leq 3\epscorr.
\end{equation}

For user privacy, we look at the total view of one data centre (say $\db_1$) together with the eavesdropper, $\eve$. However, it is not straightforward to compare the views for different $x$. Hence, we introduce an hypothetical scenario which uses an ideal QKD protocol instead of a real QKD protocol to generate keys for OTP encryption through $\channel_{U2}$ as an intermediate step. This state, $\xi_{\db_1\eve}^x=\xi_{\tilde{Q}_1\bar{A}_1S_1S_5WC_{Q_1}C_{Q_2}C_{\bar{A}_1}C_{\bar{A}_2}E}^x$ has the same set of variables as $\rho_{\db_1\eve}^x$ in Eq.~\eqref{dc1eveview}, with the only difference being that QKD keys $S_3S_4$ are ideal. With this intermediate state, we can split the trace distance into three parts by using triangle inequality, $\Delta(\rho_{\db_1\eve}^x,\xi_{\db_1\eve}^x)$, $\Delta(\xi_{\db_1\eve}^x,\xi_{\db_1\eve}^{x'})$, and $\Delta(\xi_{\db_1\eve}^{x'},\rho_{\db_1\eve}^{x'})$.

We first examine the second part, $\Delta(\xi_{\db_1\eve}^x,\xi_{\db_1\eve}^{x'})$. When the protocol aborts, the two views are clearly identical (i.e. zero trace distance) since all variables have value $\perp$, except the keys $S_1S_5E$, which are common for both states. In fact, for all trace distances we examine in this sketch proof, the two states in the trace distance are identical when the protocol aborts, and thus we ignore the protocol abort situation. When the protocol does not abort, we can simplify by using the fact that any trace-preserving map cannot increase trace distance, and noting that there are trace-preserving maps from $Q_1S_1S_2S_5W$ to $\tilde{Q}_1\bar{A}_1C_{Q_1}C_{\bar{A}_1}$. Moreover, since the ciphertext $C_{Q_2}C_{\bar{A}_2}$ is obtained from encryption using ideal QKD keys $S_3S_4$, they are uniformly distributed over $\vars{C}_{Q_2}\vars{C}_{A_2}$, and thus are independent of $x$ and common to both states. After simplification, the only remaining variable in the trace distance possibly dependent on $x$ is $Q_1$ (the other remaining variables are $S_1S_2S_5WE$). However, by 0-user privacy of the SPIR protocol, $Q_1$ is independent of $x$ and thus $\Delta(\xi_{\db_1\eve}^x,\xi_{\db_1\eve}^{x'})=0$. 

The second step is to look at the trace distance $\Delta(\rho_{\db_1\eve}^x,\xi_{\db_1\eve}^x)$. Conditioned on protocol not aborting, we can simplify by noting that there are trace-preserving maps that can map $Q_1Q_2S_1S_2S_3S_4S_5S_6W$ to $\tilde{Q}_1\bar{A}_1C_{Q_1}C_{Q_2}C_{Q_2}C_{\bar{A}_2}$. Since $Q_1Q_2$ are independent of the QKD keys, and $S_1S_2S_5S_6$ are generated by same QKD protocol, we are left with the trace distance

\be
\Delta(\rho_{\db_1\eve}^{x},\xi_{\db_1\eve}^{x})\leq(1-p_{\fail})\Delta(\rho_{S_3S_4E'}^{\top},\xi_{S_3S_4E'}^{\top}).
\ee
where $\top$ labels the conditioning on the protocol not aborting.
In the R.H.S. of the equation, the first state (resp. second state) corresponds to real QKD keys (resp. ideal QKD keys) $S_3S_4$ with side information $E'=S_1S_2S_5S_6E$ conditioned on the protocol not aborting. Therefore, from the security definition, the trace distance is bounded by $\epscorr+\epssec$. Combining the above results, one can show that

\be
\Delta(\rho_{\db_1\eve}^{x},\rho_{\db_1\eve}^{x'})\leq 2(\epscorr+\epssec).
\ee
This also holds for the total view of $\db_2\eve$.

For database privacy, we examine the total view of the user, $\user$, together with the eavesdropper Eve, $\eve$. We then introduce a hypothetical scenario where ideal QKD keys are used instead of real QKD keys as the shared random string between the data centres. The corresponding state, $\xi_{\user\eve}^{w}=\xi_{XR\bar{Q}_1\bar{Q}_2\tilde{A}_1\tilde{A}_2S_2S_4C_{\bar{Q}_1}C_{\bar{Q}_2}C_{A_1}C_{A_2}E}^{w}$, contains the same variables as $\rho_{\user\eve}$ in Eq.~\eqref{usereveview}, except that $S_5S_6$ are ideal QKD keys. Therefore, we can use triangle inequality to split the trace distance into three parts, $\Delta(\rho_{\user\eve}^{w},\xi_{\user\eve}^{w})$, $\Delta(\xi_{\user\eve}^{w},\xi_{\user\eve}^{w'})$, and $\Delta(\xi_{\user\eve}^{w'},\rho_{\user\eve}^{w'})$.

We first examine the second part, $\Delta(\xi_{\user\eve}^{w},\xi_{\user\eve}^{w'})$ for an arbitrary $x$, $r$ and ($s_1^{dec}$,$s_2^{enc}$,$s_3^{dec}$,$s_4^{enc}$). This can be simplified by noting that there is a trace-preserving map from $\bar{Q}_1\bar{Q}_2A_1A_2S_1S_2S_3S_4$ to $\tilde{A}_1\tilde{A}_2C_{\bar{Q}_1}C_{\bar{Q}_2}C_{A_1}C_{A_2}$.
Since a fixed $r$ and $x$ fixes $\bar{q}_1$ and $\bar{q}_2$ and having fixed keys ($s_1^{dec}$,$s_2^{enc}$,$s_3^{dec}$,$s_4^{enc}$) further fixes the query received by the database, $\tilde{\bar{q}}_1$ and $\tilde{\bar{q}}_2$, we can express the state as two subsystems $XR\bar{Q}_1\bar{Q}_2S_1S_2S_3S_4E$ and $\tilde{\bar{Q}}_1\tilde{\bar{Q}}_2A_1A_2$. The former subsystem is independent of $W$, and thus we can remove it using the fact that $\Delta(A\otimes B,A\otimes B)\leq \Delta(B,C)$.
The probability distribution of $\tilde{\bar{Q}}_1\tilde{\bar{Q}}_2A_1A_2$ here is the same as a hypothetical scenario where all QKD keys are ideal, and the user sends the queries $\tilde{\bar{Q}}_1$ and $\tilde{\bar{Q}}_2$ instead. For this scenario, we can invoke 0-database privacy, which states there exists an $x'$ such that for $w$ and $w'$ with $w_{x'}=w'_{x'}$, $A_1$ and $A_2$ are independent on $W$ (i.e. trace distance is zero). This is true for any adversarial user queries, and in particular it is true for queries $\tilde{\bar{Q}}_1$ and $\tilde{\bar{Q}}_2$.

The next step is to examine the trace distance $\Delta(\rho_{\user\eve}^{w},\xi_{\user\eve}^{w})$. We note that there are trace-preserving maps that can be applied to $\bar{Q}_1\bar{Q}_2S_1S_2S_3S_4S_5S_6W$ to obtain $A_1A_2C_{\bar{Q}_1}C_{\bar{Q}_2}C_{A_1}C_{A_2}$. This simplification, together with removal of common terms $XR\bar{Q}_1\bar{Q}_2W$, and noting that $S_1S_2S_3S_4$ is generated by the same QKD protocol for both terms, we arrive at

\be
\Delta(\rho_{\user\eve}^w,\xi_{\user\eve}^w)\leq (1-p_{fail})\Delta(\rho_{S_5S_6E'}^{\top},\xi_{S_5S_6E'}^{\top}),
\ee
where the side-information is $E'=S_1S_2S_3S_4E$. The terms in the trace distance corresponds to the output state of a real and ideal QKD protocol respectively conditioned on not aborting. Therefore, from the security definition, this is bounded by $\epscorr+\epssec$. Combining the above results, we conclude that there exist a $x'$ such that for $w_{x'}=w'_{x'}$,

\be
\Delta(\rho_{\user\eve}^{w},\rho_{\user\eve}^{w'})\leq 2(\epscorr+\epssec).
\ee

The final condition of protocol secrecy requires the introduction of the view of the eavesdropper for two different scenarios. $\xi_{\eve}^{x,w,1}$ is Eve's view in a setup where $S_1S_2$ are ideal QKD keys, and $\xi_{\eve}^{x,w,2}$ is Eve's view where $S_1S_2S_3S_4$ are ideal QKD keys. Using similar arguments from the sketch proof of user privacy, one can show that each change from $\rho_{\eve}^{x,w}\rightarrow\xi_{\eve}^{x,w,1}\rightarrow\xi_{\eve}^{x,w,2}$ incurs an error of $\varepsilon$, resulting in trace distance $\Delta(\rho_{\eve}^{x,w},\xi_{\eve}^{x,w,2})\leq 2(\epscorr+\epssec)$. 

The next step is to examine the trace distance $\Delta(\xi_{\eve}^{x,w,2},\xi_{\eve}^{x',w',2})$. We note that $\xi_{\eve}^{x,w,2}=\xi_{C_{Q_1}C_{Q_2}C_{A_1}C_{A_2}E}^{x,w,2}$ is similar to $\rho_{\eve}^{x,w}$ in Eq.~\eqref{eveview}, except that $S_1S_2S_3S_4$ are ideal QKD keys. Since $C_{Q_1}C_{Q_2}C_{A_1}C_{A_2}$ are ciphertext generated using ideal QKD keys $S_1S_2S_3S_4$, they are distributed uniformly over $\vars{C}_{Q_1}\vars{C}_{Q_2}\vars{C}_{A_1}\vars{C}_{A_2}$. Therefore, they are not dependent on $x$ or $w$ (neither is $E$), and the trace distance is $\Delta(\xi_{\eve}^{x,w,2},\xi_{\eve}^{x',w',2})=0$. Using triangle inequality to combine the result, we have

\be
\Delta(\rho_{\eve}^{x,w},\rho_{\eve}^{x',w'})\leq 4(\epscorr+\epssec).
\ee
The detailed proof is provided in Appendix~\ref{SecProofApp}.


\section{Numerical Simulation}
\label{NumSim}

\subsection{MDI-QKD}

\begin{figure}[H]
    \centering
    \includegraphics[width=0.7\textwidth]{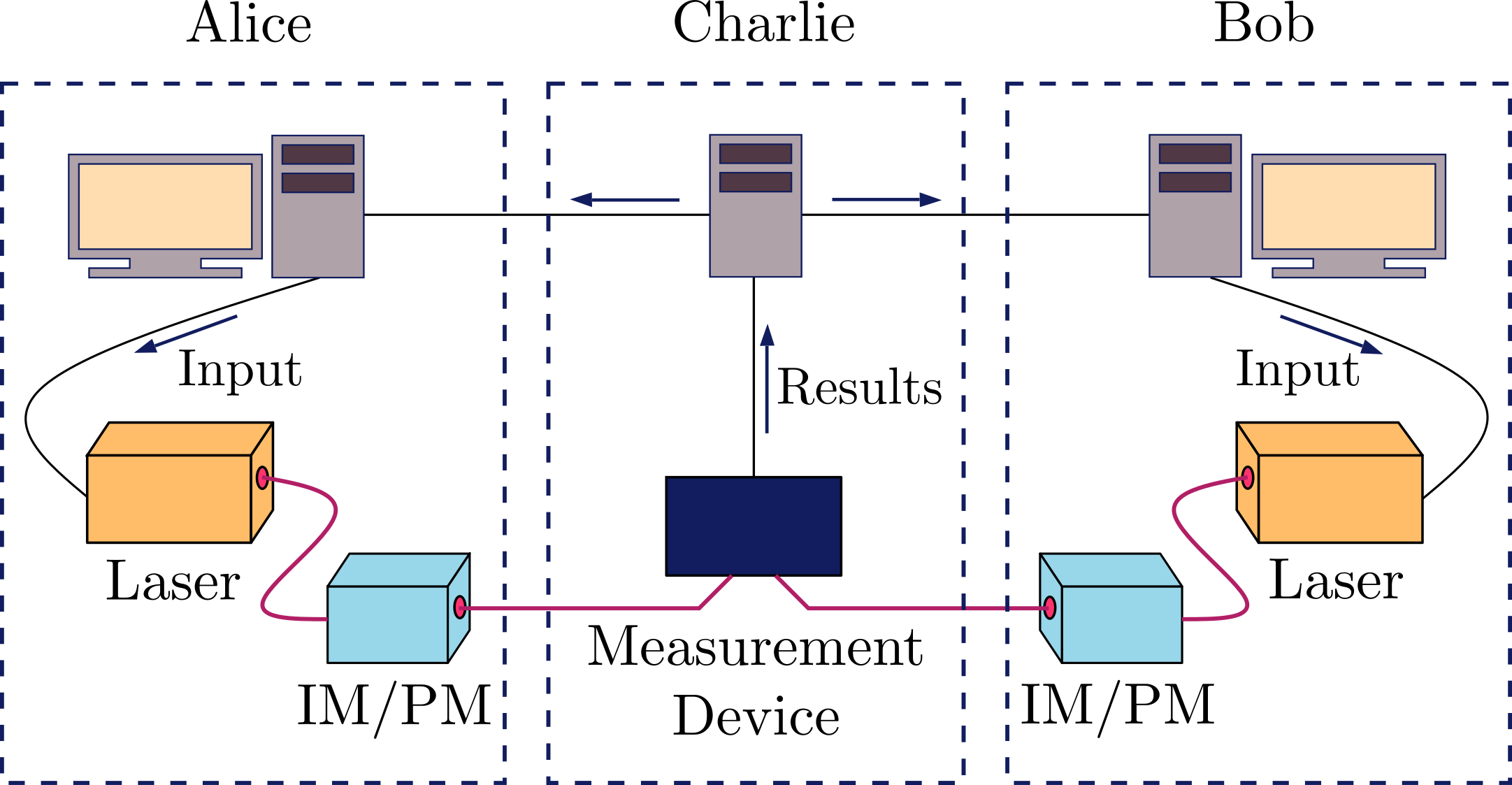}
    \caption{Setup for implementing MDI-QKD. Alice and Bob each holds a source and modulators which can be used to generate quantum states in basis X or Z and for different intensities. These states are sent to the central node (Charlie) which performs a measurement and broadcasts the result. An honest Charlie would performs Bell state measurement.}
    \label{MDIQKDFig}
\end{figure}

For simulation purposes, we look at MDI-QKD with decoy states~\cite{Curty2014} as the protocol of choice to generate the keys used in SPIR. In MDI-QKD, the security of the QKD key generated is guaranteed even if the eavesdropper is the one performing the measurement and announcing the result, as shown in Figure~\ref{MDIQKDFig}. Hence, in the setup depicted in Figure~\ref{SetupFig}, the central node would hold the measurement device and the other parties would hold the QKD source. In this case, the MDI nature of the protocol ensures that the central node cannot gain any information about the messages communicated between the user and data centres. 

The MDI-QKD protocol we use is detailed in Ref.~\cite{Curty2014}, and we provide a summary here. We start with the communicating parties, Alice and Bob, each choosing a basis from $\{X,Z\}$, an intensity from $\{a_s,a_1,\dots,a_n\}$ and $\{b_s,b_1,\dots,b_m\}$ respectively, and a random bit $\{0,1\}$. They then prepare the corresponding quantum state and send it to the central node. If the central node is honest, it will perform a Bell state measurement and report the result, $t$. Alice and Bob can then reveal their basis and intensity settings and only select rounds where they use the same basis states. This sifted key can then be used for parameter estimation, error correction and privacy amplification. The final key rate obtained is given by the sum of key rates for different results reported by the central node, $l=\sum_t l_t$, 

\begin{equation}
    \begin{split}
        l_t\leq& n_{t,0}+n_{t,1}[1-h(e_{t,1})]-\text{leak}_{\rm{EC},t}\\
        &-\log\frac{8}{\epscorr}-2\log\frac{2}{\varepsilon_t'\hat{\varepsilon}_t}-2\log\frac{1}{2\varepsilon_{t,\rm{PA}}},
    \end{split}
\end{equation}
where $h(x)$ is the binary entropy of $x$, $n_{t,0}$ is the number of events where either party sends zero photons, $n_{t,1}$ is the number of events where both parties send one photon each, $e_{t,1}$ is the error rate for these one-photon events, $\text{leak}_{\rm{EC},t}$ is the number of leaked bits from error-correction, and the $\varepsilon$ values are various security and parameter estimation parameters.

\subsection{SPIR Resource}

We examine the performance of the SPIR protocol based on the type of database it can send for a fixed number of signals sent to establish the QKD key, $N$, and for fixed distances, $d$. A database is characterised by the number of entry it has, $n$, and the size of each entry, $L$. We use the two-database SPIR protocol $\btwoprot$~\cite{Gertner2000} (see Appendix~\ref{B2ProtApp} for protocol description), which requires communication of $[7L+3\log(n^{1/3})+(3+3L)n^{1/3}]$ bits between the user and each data centre, and $(9Ln^{1/3}+10L)$ bits of shared key between the data centres for CDS. In a typical implementation, it is likely that the two data centres would be close together, thus the limiting factor would be from the user-data centre communication since the user would tend to be far from the data centre itself. Hence, we will only focus on the the key rate from MDI-QKD between the user and data centres.

\begin{figure}[H]
    \centering
    \includegraphics[width=0.6\textwidth]{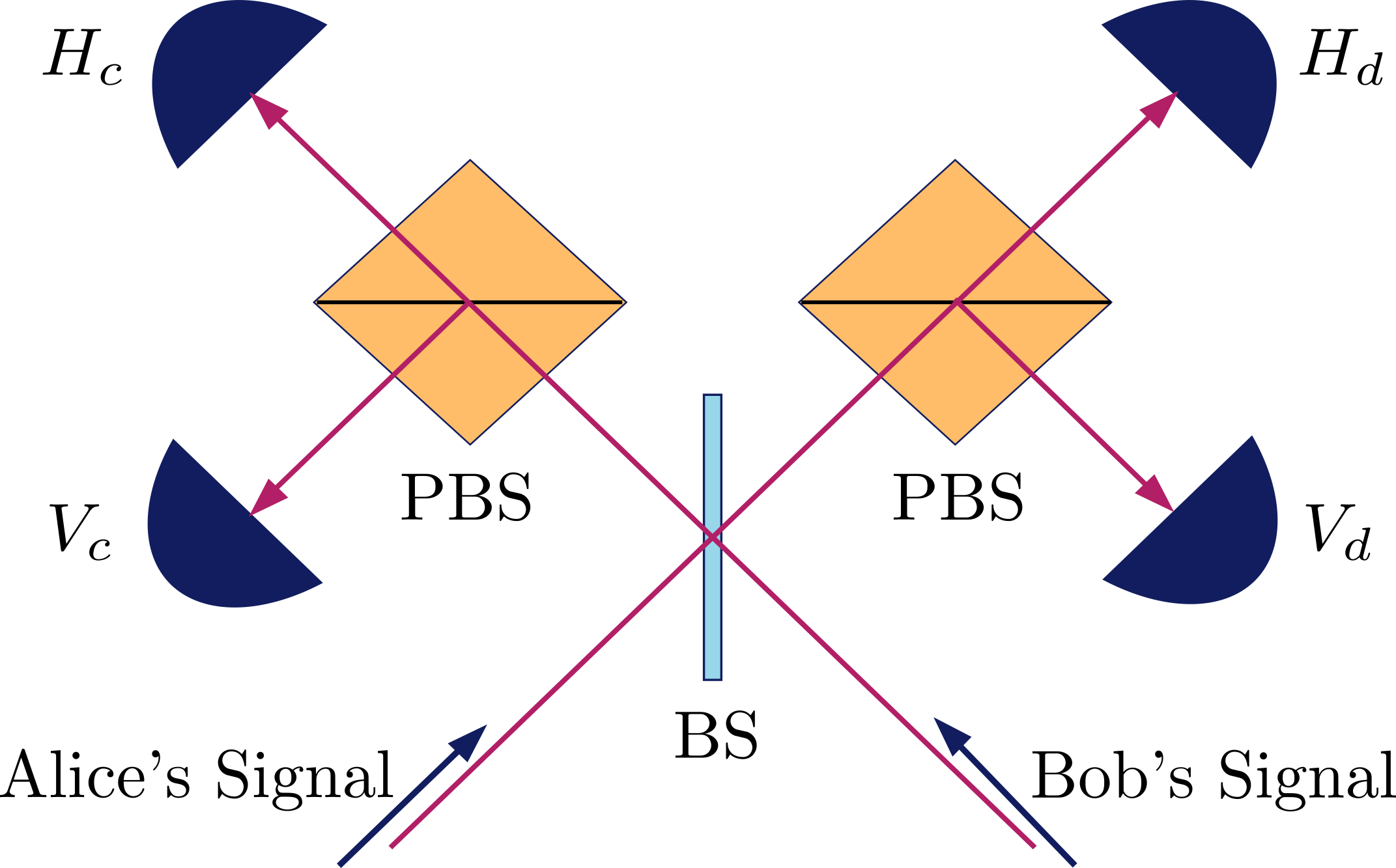}
    \caption{Schematic of measurement device held by central node. Signals sent from Alice and Bob would enter the beam splitter (BS) before being sent to two polarising beam splitters (PBS) and triggering the single photon detectors. The detectors are labelled based on the polarisation of photon and path they detect. A detection of both $H_c$ and $V_d$ or $H_d$ and $V_c$ indicates a projection to $\ket{\psi^-}$ and a detection of both $H_c$ and $V_c$ or $H_d$ and $V_d$ indicates a projection to $\ket{\psi^+}$.}
    \label{MDeviceFig}
\end{figure}

In the analysis, we use similar parameters as in Ref.~\cite{Curty2014}, with a fibre channel loss of \SI{0.2}{\decibel\per\kilo\metre}, detection efficiency of \SI{14.5}{\percent}, and background count of \SI{6.02e-6}{}. We assume that the central node uses the measurement device shown in Figure~\ref{MDeviceFig}, which allows it to perform Bell state measurements of states $\ket{\psi^-}$ and $\ket{\psi^+}$. The polarisation misalignment error of this setup is modelled following Ref.~\cite{Xu2013}, by introducing unitary rotations in the channels connecting Alice and Bob to the central node, and a unitary rotation in one arm of the measurement device after the beam splitter. The value of the total polarisation misalignment error is set at \SI{1.5}{\percent}. For simplicity, the protocol uses only two decoy states, with the weaker one having intensity \SI{5e-4}{}. We also assume that the error correction leakage is given by $\text{leak}_{\rm{EC},t}=1.16n_th(e_t^{a_sb_s})$, where $n_k$ is the number of bits of the sifted key (runs that both Alice and Bob prepares in the Z-basis and using the signal intensity) that is not used for error estimation, and $e_t^{a_sb_s}$ is the corresponding error rate of this sifted key.

\begin{figure}[H]
    \centering
    \includegraphics[width=0.95\textwidth]{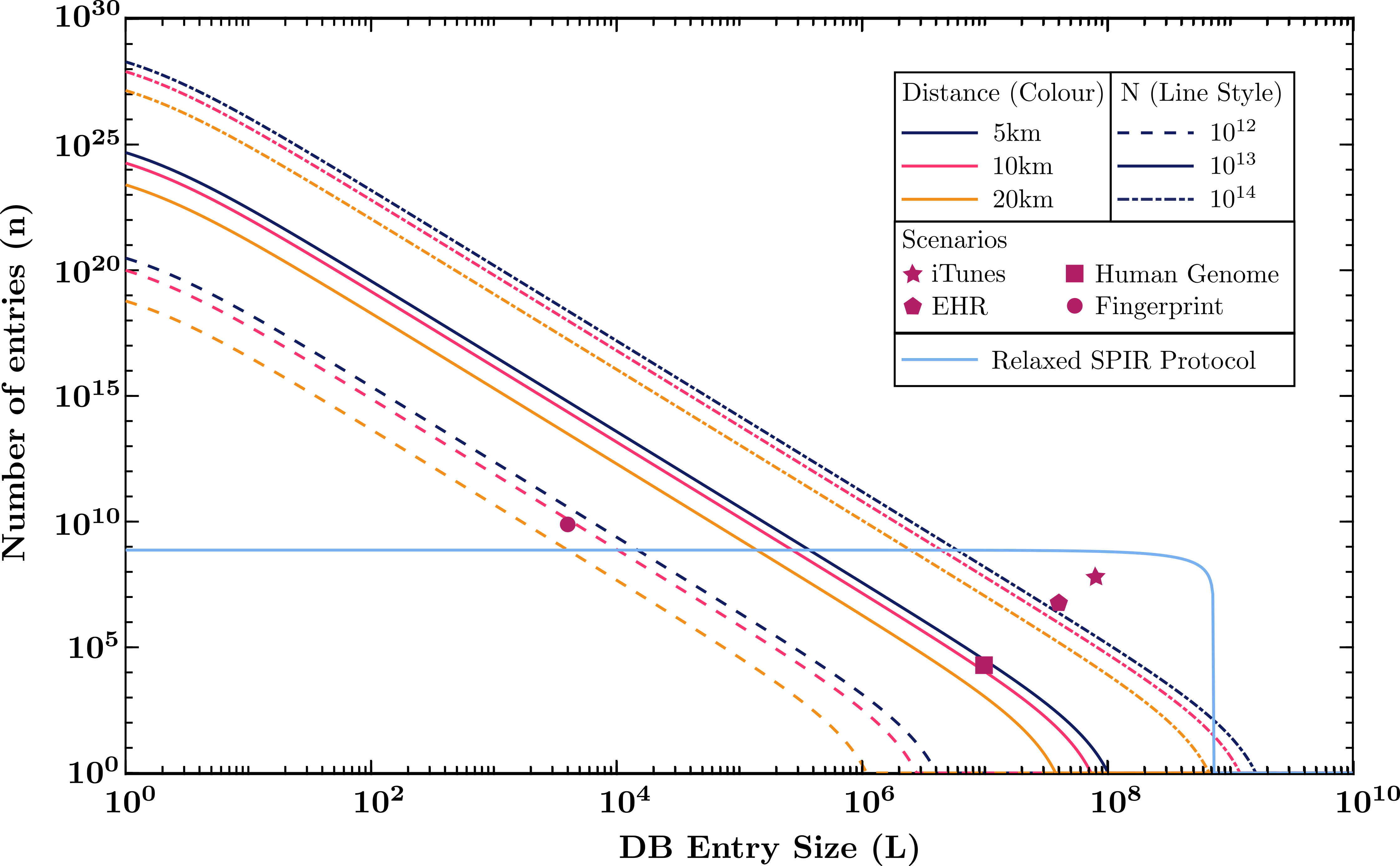}
    \caption{Plot of database parameters, $L$, the size of each entry of the database, and $n$, the number of entries in the database for various number of signals sent, $N$, (labelled by different line style) and distances, $d$ (labelled by different colours). Four points are included that represents the database parameters of the usage scenarios described in the main text. The diagram also includes a plot for an alternative protocol that requires a more relaxed SPIR definition discussed in Sec.~\ref{SecDiscuss}.}
    \label{PlotFig}
\end{figure}

We fix the QKD security parameters $\varepsilon_{corr}=$\SI{e-15}{} and $\varepsilon=$\SI{e-10}{}, which makes the SPIR (\SI{3e-15}{}, \SI{2e-10}{}, \SI{2e-10}{}, \SI{4e-10}{})-secure. The key rate $l/N$ is optimised for a given number of signals sent in the QKD key generation, $N$, over all free parameters. These include the intensities, probability distributions of intensity and basis choices, number of bits used for error estimation, and the security parameters implicit in $\varepsilon$. We plot the database parameters for a few setups, with the number of signal sent, $N$, being \SI{e12}{}, \SI{e13}{}, and \SI{e14}{}, which corresponds to \SI{16.7}{\minute}, \SI{2.8}{\hour}, and \SI{28}{\hour} respectively for a \SI{1}{\giga\hertz} signal rate. The distances used are metropolitan, at \SI{5}{\kilo\metre} (fits Singapore's downtown core), \SI{10}{\kilo\metre} (fits Geneva, London inner ring road), and \SI{20}{\kilo\metre} (fits Washington DC). We also included four scenarios of database query usage,
\begin{itemize}[leftmargin=*,labelsep=5.8mm]
    \item iTunes: A consumer wants to purchase a song from the iTunes catalogue, which contains 60 million songs. (Assume each music file is 10MB) [$n=$ \SI{6e7}{}, $L=$ \SI{8e7}{}]
    \item Electronic Health Records (EHR): A doctor in Singapore wants to retrieve his patient's medical chart from the national health records database. (The average medical chart file size of a healthy patient is about 5MB~\cite{FCC2010}, and Singapore's population is 5.7 million~\cite{Singapore_Pop}) [$n=$ \SI{5.7e6}{}, $L=$ \SI{4e7}{}]
    \item Fingerprint Data: Border control wants to retrieve the fingerprint data of a visitor to verify his identity. (Fingerprint minutiae data size is about 500 bytes~\cite{ISO_Fingerprint}, and the world population is 7.7 billion~\cite{World_Pop}) [$n=$ \SI{7.7e9}{}, $L=$ \SI{4000}{}]
    \item Genetic Data: A doctor requests for a gene in a patient's genome data to analyse disease risk. (Human genome contains 19116 protein-coding genes, with the maximum size of a single gene being 2.47 million base pairs~\cite{Human_Genome}. Since humans have two alleles for most genes and there are 4 possible bases, each gene entry can be encoded as 9.88 million bits). [$n=$ \SI{19116}{}, $L=$ \SI{9.88e6}{}]
\end{itemize}
The results are shown in Figure~\ref{PlotFig}.

The $\btwoprot$ protocol with QKD has a scaling of $O(n^{1/3}L)$, which is reflected in the numerical analysis by the significantly higher number of database entries that one can perform SPIR for compared to the database entry size, which scales linearly with $N$. This means that the $\btwoprot$ protocol is especially useful for databases with small file sizes and large number of entries, such as querying the fingerprint of one person from a database containing the fingerprint of everyone in the world, which takes about 16.7 minutes of key generation for \SI{10}{\kilo\metre} distances. For much larger database entries, such as video files, and uncompressed music files, the use of the $\btwoprot$ protocol with QKD channels does not appear feasible. 

\section{Discussion}
\label{SecDiscuss}

Having a multi-database SPIR protocol with QKD provides information theoretic security, but a drawback in the setup is that the result obtained by the user, $\hat{w}_x$, cannot be verified. This allows malicious data centres to send false information to the user simply by changing the answers sent to the user. This, however, does not affect the validity of the SPIR protocol. At the practical level, this act could be detectable for certain applications, such as music streaming, but could remain undetected for other applications such as medical test reports, where information cannot be independently verified by the user. One could overcome this by providing additional information, such as a hash of the desired entry, for the user to perform verification, but this requires a further analysis which is beyond the scope of the current work.

In place of ideal keys, we have introduced the use of QKD keys for use in SPIR, but we require a few addition assumptions on the parties. In particular, we assume that (1) the data centres do not intentionally leak the QKD keys to other parties including Eve, (2) that all messages sent through the channels $\channel_{Uj}$ must be encrypted with OTP, and that (3) data centres do not have access to the classical channels used to establish the QKD keys after the key exchange step. These additional assumptions are necessary to prevent the misuse of QKD, which assumes that both communicating parties act honestly. These assumptions can be enforced in practice via methods like supervisory programs or a trusted third party authority. 

In our numerical analysis, we used the $\btwoprot$ protocol, but there are other SPIR protocol that one could use. $\mathscr{B}_k''$ protocol is a generalisation of the $\btwoprot$ protocol that requires $k$ databases instead of a two, with a scaling of $O(n^{1/(2k-1)}L)$. This means that it outperforms the $\btwoprot$ for applications with a large number of database entries, but the user would have to communicate with more data centres. 

Alternatively, one could relax the SPIR definition to allow for other protocols to be used. In the current SPIR definition, the user is not allowed to learn the values of the XOR of database entries such as $w_x\oplus w_{x'}$. However, in certain scenarios the data centre might not mind the user learning such values, as long as the user only gains one bit of information, e.g., either $w_x$ or some $\bigoplus_x w_x$. Such a change would require further modification of Definition~\ref{DBPrivDefn}, for instance, to one that reads ``there exist an $i^{(n)}=(i_1,\dots,i_n)$ such that for all $w$ and $w'$ with $\bigoplus_x i_xw_x=\bigoplus_x i_xw_x'$", where $i_x=1$ indicates that the user includes $w_x$ in the XOR the user learns and $i_x=0$ otherwise. 

The relaxation made to the SPIR definition would allow us to use another protocol used as the foundation for Song et. al.'s quantum SPIR protocol~\cite{Song2019}. In this protocol, we label the user's desired bit as $w_{i^{(n)}}=\bigoplus_{x=1}^{n}i_x w_{x}$. The user then generates a random string $R^{(n)}\in\{0,1\}^n$ and sends his queries $Q_1^{(n)}=R^{(n)}$, $Q_2^{(n)}=R^{(n)}\oplus i^{(n)}$. The data centres then reply with answers $A_1=\left(\bigoplus_{x=1}^n Q_{1,x}w_x\right)\oplus K$ and $A_2=\left(\bigoplus_{x=1}^n Q_{2,x}w_x\right)\oplus K$, where $K$ is a shared random bit between the data centres. The user would then decode by applying $A_1\oplus A_2$, and $K$ ensures that the user can only obtain at most a single bit. In this setup, the number of bits of communication between the user and data centre is $n+L$, and the plot is shown in Figure~\ref{PlotFig}, for $N=$ \SI{e13} at 10km. This protocol can be utilised for iTunes and EHR, which is not feasible for the $\btwoprot$ protocol. The protocol can also achieve close to the communication limit of $L=l$ for small databases. This limit is that of the secure communication of a single string (entry) of length $L$, which requires one QKD secure key bit for each bit of the string. However, the number of entries that the database can have is limited in this case, and it can no longer be used for the fingerprint database which has 7.7 billion entries. Therefore, it can be useful to examine other protocols of SPIR or relaxed versions of SPIR.

Here, we have shown how multi-database SPIR can work with QKD channels in place of secure channels. An interesting extension would be to demonstrate it experimentally, which would pave the way for practical implementation of the protocol in the future. For practical implementation, it is also useful to explore reasonable relaxations of the QKD protocol, such as the one described above, and other SPIR or relaxed SPIR protocols. By optimising the protocol choice for different applications of SPIR based on the number of entries and database entry size of the database, one could obtain better performance for the particular application of interest. 

Another interesting extension would be to examine the performance of SPIR in the situation of a byzantine adversary who may corrupt transmission for some of the communication channels, and the scenario where this adversary can collude with some data centres. This situation results in communication between the data centres, which could compromise user privacy, and inaccurate answers being sent to the user due to corrupted transmission, which could affect the correctness of the protocol. The classical case was examined by Wang et. al.~\cite{Wang2017}, where they also looked at the scenario where an eavesdropper that can tap into the communication channels, but this problem has been addressed in this paper with QKD. It is thus interesting to explore if the quantum nature of the byzantine adversary and the colluding data centres could have an impact on SPIR implementation with QKD channels. The SPIR solution to this scenario would provide additional security for the user.


\section{Conclusion}

We have introduced the use of QKD in place of secure channels in SPIR, since classical secure channels are difficult to achieve in practice. To show that replacing the classical secure channel with QKD channels does not compromise security, we extended the original SPIR definition to include aspects of QKD that are not normally present in a secure channel. These include the presence of an external eavesdropper who may tap into classical communication and eavesdrop on the quantum channel, having security parameters due to the possibility of having an imperfect secret key and considering that the QKD protocol may abort. Using the extended SPIR definition, we then show that the SPIR security parameters are related to the QKD security parameters, $\epssec$ and $\epscorr$, which can be set arbitrarily close to zero. This implies that one could have a SPIR protocol using QKD keys with arbitrarily good security. Using MDI-QKD and $\btwoprot$ protocol as an example, we also show how such a SPIR protocol, specifically $\btwoprot$, can be feasible by numerically simulating the QKD key rates. 

\vspace{6pt} 

\authorcontributions{Conceptualization, C.C.W.L.; Formal analysis, W.Y.K. and C.C.W.L.; Investigation, W.Y.K.; Visualization, W.Y.K.; Writing--original draft and review \& editing, W.Y.K. and C.C.W.L.; Supervision, C.C.W.L.; Funding acquisition, C.C.W.L..}


\funding{This research was funded by the National Research Foundation of Singapore: NRF Fellowship grant (NRFF11-2019-0001) and NRF Quantum Engineering Programme grant (QEP-P2). W.Y. Kon acknowledges support from the NUS President's Graduate Fellowship (funded by Lee Kong Chian Scholarship).}

\acknowledgments{We thank Chao Wang, Ignatius William Primaatmaja, and Koon Tong Goh for their comments and useful suggestions. We also thank the referees from the Quantum Journal for their constructive comments.}


\conflictsofinterest{The authors declare no conflict of interest.}

\abbreviations{The following abbreviations are used in this manuscript:\\

\noindent 
\begin{tabular}{@{}ll}
PIR & Private information retrieval\\
SPIR & Symmetric private information retrieval\\
QKD & Quantum key distribution\\
CPTP & Completely positive and trace preserving\\
POVM & Positive operator value measurement\\
OTP & One-time pad\\
CDS & Conditional disclosure of secrets\\
MDI & Measurement-device independent
\end{tabular}}

\appendixtitles{yes} 
\appendix

\section{Detailed Security Proof}
\label{SecProofApp}

\begin{Theorem}
    A two-database one-round \textnormal{(0,0,0,0)}-secure SPIR protocol that uses $\varepsilon$-secure QKD keys in place of ideal keys, where $\varepsilon=\epscorr+\epssec$, is $3\epscorr$-correct.
\end{Theorem}
\begin{proof}

We start by noting that when all the QKD keys are correct, $S_1=S_2$, $S_3=S_4$, and $S_5=S_6$, answers generated by the data centres and messages sent through the channels would be correct. From the 0-correctness of the classical SPIR protocol, this means that the user would be able to decode correctly, $\hat{w}_x=w_x$. Therefore, we have the result in Eq.~\eqref{CorrEqn}. Taking the complement of Eq.~\eqref{CorrEqn} gives

\begin{multline}
        \Pr[\hat{w}_x\neq w_x|\pass]
        \leq \Pr \left[\{ S_1\neq S_2\}\cup \{S_3\neq S_4\}
        \cup\{S_5\neq S_6\}|\pass \right]
        \\\leq \Pr\left[S_1 \neq S_2|\pass\right]+ \Pr\left[S_3\neq S_4|\pass \right] 
         +\Pr\left[S_5\neq S_6|\pass\right],
\end{multline}
where the second inequality is an application of the union bound. This can be directly related to $\epscorr$ of each channel to give the correctness condition,

\begin{multline}
    (1-p_{\fail})\Pr[\hat{w_x}\neq w_x|\pass]
    \\\leq (1-p_{\fail})\left(\frac{\epscorr^{U1}}{1-p_{\abort,U1}}+\frac{\epscorr^{U2}}{1-p_{\abort,U2}}\right.
    \left.+\frac{\epscorr^{12}}{1-p_{\abort,12}}\right)\\
    \leq \epscorr^{U1}+\epscorr^{U2}+\epscorr^{12}=3\epscorr,
\end{multline}
where the second inequality is obtained noting that the probability that the SPIR protocol would abort, $p_{\fail}$, is larger than the probability that any one QKD protocol aborts, $p_{\perp}$.
\end{proof}

\begin{Theorem}
    A two-database one-round \textnormal{(0,0,0,0)}-secure SPIR protocol that uses $\varepsilon$-secure QKD keys in place of ideal keys, where $\varepsilon=\epscorr+\epssec$, is $2\varepsilon$-user private.
\end{Theorem}

\begin{proof} 
Here, we only provide the security analysis with respect to data centre 1, which can act dishonestly; the same procedure holds for data centre 2.
To compare the total view of $\db_1$ and $\eve$ for different user desired index, $\rho_{\db_1\eve}^x$ and $\rho_{\db_1\eve}^{x'}$, we first have to introduce an intermediate state, $\xi_{\db_1\eve}^{x}$. 
This state corresponds to a setup in which an ideal QKD key is generated from the QKD protocol for communication between $\db_2$ and $\user$. Using the triangle inequality property of the trace distance measure, we split the user privacy condition into three parts,

\begin{equation}\label{up_proof_eq1}
    \Delta(\rho_{\db_1\eve}^x,\rho_{\db_1\eve}^{x'})\leq \Delta(\rho_{\db_1\eve}^x,\xi_{\db_1\eve}^x)
    +\Delta(\xi_{\db_1\eve}^x,\xi_{\db_1\eve}^{x'})
    +\Delta(\xi_{\db_1\eve}^{x'},\rho_{\db_1\eve}^{x'}).
\end{equation}

We start by examining the second term on the R.H.S., which is the trace distance between two views where the secret key pairs used are $(S_1,S_2)$ and $(S_5,S_6)$ from the actual QKD protocols, and $(S_3,S_4)$ from an ideal QKD protocol, but with differing user index choices $x$ and $x'$. Following Eq.~(\ref{dc1eveview}), we have that

\begin{multline}
\xi_{\db_1\eve}^x=\xi_{\tilde{Q}_1\bar{A}_1S_1S_5WC_{Q_1}C_{Q_2}C_{\bar{A}_1}C_{\bar{A}_2}E}^x\\
=(1-p_{\fail})\xi_{\tilde{Q}_1\bar{A}_1S_1S_5WC_{Q_1}C_{Q_2}C_{\bar{A}_1}C_{\bar{A}_2}E}^{x,\top}+p_{\fail}\xi_{\tilde{Q}_1\bar{A}_1S_1S_5WC_{Q_1}C_{Q_2}C_{\bar{A}_1}C_{\bar{A}_2}E}^{x,\perp}
\end{multline}
where the label $\top$ indicates that the state is conditioned on the QKD not aborting (i.e. All QKD keys are not $\perp$) and $\perp$ indicates that the state is conditioned on QKD aborting. We note that the state conditioned on aborting would have all terms being $\perp$ except possibly the QKD keys and $W$. Therefore, it is clear that this is independent of $X$,

\be
\Delta\left(\xi_{\tilde{Q}_1\bar{A}_1S_1S_5WC_{Q_1}C_{Q_2}C_{\bar{A}_1}C_{\bar{A}_2}E}^{x,\perp},\xi_{\tilde{Q}_1\bar{A}_1S_1S_5WC_{Q_1}C_{Q_2}C_{\bar{A}_1}C_{\bar{A}_2}E}^{x',\perp}\right)
=\Delta\left(\xi_{S_1S_5WE}^{x,\perp},\xi_{S_1S_5WE}^{x',\perp}\right)
=0
\ee
Then, by noting the following trace-preserving mappings

\begin{multline*}
Q_1S_2 \overset{{\rm{partial\, trace}}}{\mapsto} Q_1S_2^\enc \overset{\rm{OTP}}{\mapsto} C_{Q_1}, \quad  C_{Q_1}S_1 \overset{\rm{partial\, trace}}{\mapsto} S_1^{dec}C_{Q_1} \overset{\rm{OTP}}{\mapsto} \tilde{Q}_1,\\
\tilde{Q}_1WS_5 \overset{\baranswera}{\mapsto} \bar{A}_1, \quad {\rm{and}}\quad \bar{A}_1S_1 \overset{{\rm{partial\, trace}}}{\mapsto} \bar{A}_1S_1^\enc \overset{\rm{OTP}}{\mapsto} C_{\bar{A}_1},  
\end{multline*}
and using the jointly convex property of trace distance, we further get 

\be
\Delta(\xi_{\db_1\eve}^x,\xi_{\db_1\eve}^{x'})\leq (1-p_{\fail})\Delta\left(\xi_{Q_1S_1S_2S_5WC_{Q_2}C_{\bar{A}_2}E}^{x,\top},\xi_{Q_1S_1S_2S_5WC_{Q_2}C_{\bar{A}_2}E}^{x',\top}\right).
\ee
At this point, we note that $C_{Q_2}$ and $C_{A_2}$ are encrypted with an ideal secret key and hence is uniformly distributed whenever the protocol does not abort. More specifically, $C_{Q_2}$ (resp. $C_{\bar{A}_2}$) is uniformly distributed over $\vars{C}_{Q_2}$ (resp. $\vars{C}_{\bar{A}_2}$) with probability $1-p_{\fail}$. 
With this, we can expand the trace distance to get

\begin{multline}
\Delta(\xi_{\db_1\eve}^x,\xi_{\db_1\eve}^{x'})
\leq (1-p_{\fail}) 
\Delta \left( \xi_{Q_1S_1S_2S_5W}^{x,\top}\otimes \sum_{\mathclap{c_{q_2},c_{a_2}}}\frac{\Pi_{C_{Q_2}C_{\bar{A}_2}}(c_{q_2}c_{a_2})}{|\vars{C}_{Q_2}||\vars{C}_{A_2}|} \otimes \sigma_E^{s_1,s_2,s_5},\right.\\ \left.\xi_{Q_1S_1S_2S_5W}^{x',\top}\otimes \sum_{\mathclap{c_{q_2},c_{a_2}}}\frac{\Pi_{C_{Q_2}C_{\bar{A}_2}}(c_{q_2}c_{a_2})}{|\vars{C}_{Q_2}||\vars{C}_{A_2}|} \otimes \sigma_E^{s_1,s_2,s_5}\right).
\end{multline}
Note that $Q_1$ and $S_1S_2S_5WC_{Q_2}C_{\bar{A}_2}E$ are independent of each other, and that $S_1S_2S_5WC_{Q_2}C_{\bar{A}_2}E$ is independent of $X$.
In fact, $C_{Q_2}$ and $C_{\bar{A}_2}$ contains no information about $Q_2$ and $\bar{A}_2$ and thus none of $X$ as well. Thus, this gives us

\be
\Delta(\xi_{\db_1\eve}^x,\xi_{\db_1\eve}^{x'})
\leq  (1-p_{\fail})\Delta(\xi_{Q_1}^{x,\top},\xi_{Q_1}^{x',\top}) \leq (1-p_{\fail})\Delta(V_{\db_1}^x,V_{\db_1}^{x'})=0. 
\ee
The second inequality is due to the fact that $Q_1$ is diagonal, which means that the trace distance between probability distribution of $Q_1$ coincides with the quantum state, and that $Q_1$ is part of the view $V_{\db_1}^x$. Since $Q_1$ is generated by a honest user and thus independent on the type of channel used in the protocol, the last equality holds due to 0-user privacy of the classical protocol.

Let us now examine the first term on the R.H.S. of Eq.~(\ref{up_proof_eq1}), $\Delta(\rho_{\db_1\eve}^x,\xi_{\db_1\eve}^x)$. Likewise, we have that 

\be
\Delta(\rho_{\db_1\eve}^x,\xi_{\db_1\eve}^{x})\leq (1-p_{\fail})\Delta\left(\rho_{Q_1S_1S_2S_5WC_{Q_2}C_{\bar{A}_2}E}^{x,\top},\xi_{Q_1S_1S_2S_5WC_{Q_2}C_{\bar{A}_2}E}^{x,\top}\right).
\ee
Here, we note that the following trace-preserving mappings are applied to $Q_2S_3S_4S_6W$ to get $C_{Q_2}C_{\bar{A}_2}$, 

\[Q_2S_3S_4 \overset{{\rm{partial\, trace}}}{\mapsto} Q_2S_3^\dec S_4^\enc \overset{\rm{OTP}}{\mapsto} C_{Q_2}S_3^\dec \overset{\rm{OTP}}{\mapsto} \tilde{Q}_2, \quad  S_6W\tilde{Q}_2 \overset{\baranswerb}{\mapsto} \bar{A}_2,\quad \bar{A}_2S_3 \overset{{\rm{partial\, trace}}}{\mapsto} \bar{A}_2S_3^\enc \overset{\rm{OTP}}{\mapsto} C_{\bar{A}_2}. \]
Therefore, we get

\be
\Delta\left(\rho_{Q_1S_1S_2S_5WC_{Q_2}C_{\bar{A}_2}E}^{x,\top},\xi_{Q_1S_1S_2S_5WC_{Q_2}C_{\bar{A}_2}E}^{x,\top}\right) \leq \Delta\left(\rho_{Q_1Q_2S_1S_2S_3S_4S_5S_6WE}^{x,\top},\xi_{Q_1Q_2S_1S_2S_3S_4S_5S_6WE}^{x,\top}\right). 
\ee
We note that $Q_1Q_2$ are the only systems that depend on $x$ and that they are independent of $S_1S_2S_3S_4S_5S_6E$; recall that $Q_1Q_2$ are created independently after the QKD steps. Moreover, $W=w$ is fixed and is common to both states. These arguments thus gives us

\be
\Delta\left(\rho_{Q_1Q_2S_1S_2S_3S_4S_5S_6WE}^{x,\top},\xi_{Q_1Q_2S_1S_2S_3S_4S_5S_6WE}^{x,\top}\right)\leq \Delta\left(\rho_{S_1S_2S_3S_4S_5S_6E}^{\top},\xi_{S_1S_2S_3S_4S_5S_6E}^{\top}\right).  
\ee
Here, we can further partition $S_1S_2S_3S_4S_5S_6E$ into two parts, $S_3S_4$ and $S_1S_2S_5S_6E$, and note that $S_1S_2S_5S_6$ is common to both setups (generated using real QKD protocol). With this, we may view the latter as some extended side-information $E'=S_1S_2S_5S_6E$. Then, using the security definition of QKD (by replacing $E$ by $E'$), we get that

\begin{multline}
\Delta(\rho_{\db_1\eve}^x,\xi_{\db_1\eve}^x)\leq(1-p_{\fail})\Delta\left(\rho_{S_1S_2S_3S_4S_5S_6E}^{\top},\xi_{S_1S_2S_3S_4S_5S_6E}^{\top}\right)\\
=(1-p_{\fail})\Delta\left(\rho_{S_3S_4E'}^{\top},\xi_{S_3S_4E'}^{\top}\right)=\frac{1-p_{\fail}}{1-p_{\perp,U2}} \Delta\left(\rhoQKD_{S_3S_4E'},\rhoideal_{S_3S_4E'}\right)\leq \epscorr +\epssec,
\end{multline} 
since $\xi_{S_3S_4E'}^{\top}$ is an ideal QKD output state conditioned on not aborting. Combining the results, we obtain

\be
\Delta(\rho_{\db_1\eve}^x,\rho_{\db_1\eve}^{x'})\leq 2(\epscorr+\epssec).
\ee
\end{proof}

\begin{Theorem}
    A two-database one-round \textnormal{(0,0,0,0)}-secure SPIR protocol that uses $\varepsilon$-secure QKD keys in place of ideal keys, where $\varepsilon=\epscorr+\epssec$, is $2\varepsilon$-database private.
\end{Theorem}

\begin{proof}
We start the proof by fixing an arbitrary $x$ since the adversarial queries $\bar{Q}_1$ and $\bar{Q}_2$ sent by the user need not depend on $x$ in general. Similar to the analysis of user-privacy, we first introduce an intermediate view, $\xi_{\user\eve}^w$, that corresponds to a setup in which the QKD channel between the data centres generates an ideal output state. Using this state, we can then expand the trace distance in the database privacy condition using the triangle inequality,

\be \label{dp_proof_eq1}
\Delta(\rho_{\user\eve}^w,\rho_{\user\eve}^{w'})\leq  \Delta(\rho_{\user\eve}^w,\xi_{\user\eve}^w)
+\Delta(\xi_{\user\eve}^w,\xi_{\user\eve}^{w'})
+\Delta(\xi_{\user\eve}^{w'},\rho_{\user\eve}^{w'}),
\ee 
where for some $x'$ we have that $w\neq w'$ but with $w_{x'}=w'_{x'}$. 
To start with, we examine the second term on the R.H.S. From Eq.~(\ref{usereveview}), we have 

\be
\Delta(\xi_{\user\eve}^w,\xi_{\user\eve}^{w'})=\Delta\left(\xi_{XR\bar{Q}_1\bar{Q}_2\tilde{A}_1\tilde{A}_2S_2S_4C_{\bar{Q}_1}C_{\bar{Q}_2}C_{A_1}C_{A_2}E}^w,\xi_{XR\bar{Q}_1\bar{Q}_2\tilde{A}_1\tilde{A}_2S_2S_4C_{\bar{Q}_1}C_{\bar{Q}_2}C_{A_1}C_{A_2}E}^{w'}\right),
\ee
Then, given the following trace-preserving classical mappings, 

\begin{multline*}
A_1S_1S_2\overset{\rm{partial\,trace}}{\mapsto}A_1S_1^\enc S_2^\dec \overset{\rm{OTP}}{\mapsto}C_{A_1}S_2^\dec \overset{\rm{OTP}}{\mapsto} \tilde{A}_1, \quad A_2S_3S_4\overset{\rm{partial\,trace}}{\mapsto}A_2S_3^\enc S_4^\dec \overset{\rm{OTP}}{\mapsto}C_{A_2}S_4^\dec \overset{\rm{OTP}}{\mapsto} \tilde{A}_2,\\
\bar{Q}_1S_2\overset{\rm{partial\,trace}}{\mapsto}\bar{Q}_1S_2^{\enc}\overset{\rm{OTP}}{\mapsto}C_{\bar{Q}_1}, \quad \bar{Q}_2S_4\overset{\rm{partial\,trace}}{\mapsto}\bar{Q}_2S_4^{\enc}\overset{\rm{OTP}}{\mapsto}C_{\bar{Q}_2},
\end{multline*}
and using the jointly convex property of trace distance, we get 

\be
\Delta(\xi_{\user\eve}^w,\xi_{\user\eve}^{w'})\leq(1-p_{\fail})\Delta\left(\xi_{XR\bar{Q}_1\bar{Q}_2A_1A_2S_1S_2S_3S_4E}^{w,\top},\xi_{XR\bar{Q}_1\bar{Q}_2A_1A_2S_1S_2S_3S_4E}^{w',\top}\right).
\ee
We note that in the definition of database privacy, the trace distance is examined for a fixed (but arbitrary) $x$, $r$ and cryptographic keys ($s_1^{dec}$,$s_2^{enc}$,$s_3^{dec}$,$s_4^{enc}$). Hence, we can express

\begin{multline}
\xi_{XR\bar{Q}_1\bar{Q}_2A_1A_2S_1S_2S_3S_4E}^{w,\top}=\Pi_X(x)\otimes\Pi_R(r)\otimes\Pi_{\bar{Q}_1\bar{Q}_2}(\bar{q}_1\bar{q}_2)\otimes\Pi_{S_1^{dec}S_2^{enc}S_3^{dec}S_4^{enc}}(s_1^{dec}s_2^{enc}s_3^{dec}s_4^{enc})\\
\otimes \sum_{s_1^{enc}s_2^{dec}s_3^{enc}s_4^{dec}}P_{S_1^{enc}S_2^{dec}S_3^{enc}S_4^{dec}}(s_1^{enc}s_2^{dec}s_3^{enc}s_4^{dec})\Pi_{S_1^{enc}S_2^{dec}S_3^{enc}S_4^{dec}}(s_1^{enc}s_2^{dec}s_3^{enc}s_4^{dec})\otimes \sigma_E^{s_1s_2s_3s_4} \otimes \xi_{A_1A_2}^{w,\top,\tilde{\bar{q}}_1,\tilde{\bar{q}}_2},
\end{multline}
where we note that the adversarial queries $\bar{q}_1$ and $\bar{q}_2$ are fixed by $r$ and possibly $x$. Since $\tilde{\bar{Q}}_1=\bar{Q}_1\oplus S_2^{enc}\oplus S_1^{dec}$ and $\tilde{\bar{Q}}_2=\bar{Q}_2\oplus S_4^{enc}\oplus S_3^{dec}$, and given that the queries and keys are fixed, we can introduce $\Pi_{\tilde{\bar{Q}}_1\tilde{\bar{Q}}_2}(\tilde{\bar{q}}_1\tilde{\bar{q}}_2)$ into the state, giving

\begin{multline}
\Delta\left(\xi_{XR\bar{Q}_1\bar{Q}_2A_1A_2S_1S_2S_3S_4E}^{w,\top},\xi_{XR\bar{Q}_1\bar{Q}_2A_1A_2S_1S_2S_3S_4E}^{w',\top}\right)\\
\leq\Delta\left(\xi_{XR\bar{Q}_1\bar{Q}_2\tilde{\bar{Q}}_1\tilde{\bar{Q}}_2A_1A_2S_1S_2S_3S_4E}^{w,\top},\xi_{XR\bar{Q}_1\bar{Q}_2\tilde{\bar{Q}}_1\tilde{\bar{Q}}_2A_1A_2S_1S_2S_3S_4E}^{w',\top}\right).
\end{multline}
Since the subsystem $XR\bar{Q}_1\bar{Q}_2S_1S_2S_3S_4E$ is independent of $w$ and the subsystem $\tilde{\bar{Q}}_1\tilde{\bar{Q}}_2A_1A_2$ is independent on $S_1^{enc}S_2^{dec}S_3^{enc}S_4^{dec}$, we can remove $XR\bar{Q}_1\bar{Q}_2S_1S_2S_3S_4E$ using the fact that $\Delta(A\otimes B,A\otimes C)\leq \Delta(B,C)$,

\be
\Delta\left(\xi_{XR\bar{Q}_1\bar{Q}_2\tilde{\bar{Q}}_1\tilde{\bar{Q}}_2A_1A_2S_1S_2S_3S_4E}^{w,\top},\xi_{XR\bar{Q}_1\bar{Q}_2\tilde{\bar{Q}}_1\tilde{\bar{Q}}_2A_1A_2S_1S_2S_3S_4E}^{w',\top}\right)\leq \Delta\left(\xi_{\tilde{\bar{Q}}_1\tilde{\bar{Q}}_2A_1A_2}^{w,\top},\xi_{\tilde{\bar{Q}}_1\tilde{\bar{Q}}_2A_1A_2}^{w',\top}\right).
\ee
Since the answer functions are not dependent on the channel type (ideal or real QKD), we can equivalently view the system $\xi_{\tilde{\bar{Q}}_1\tilde{\bar{Q}}_2A_1A_2}^{w,\top}$ as one where there are ideal keys. In this case, the user sends the adversarial queries $\tilde{\bar{Q}}_1$ and $\tilde{\bar{Q}}_2$, and receives the corresponding answer $A_1$ and $A_2$. Therefore, there exist a $x'$ such that

\be
\Delta\left(\xi_{\tilde{\bar{Q}}_1\tilde{\bar{Q}}_2A_1A_2}^{w,\top},\xi_{\tilde{\bar{Q}}_1\tilde{\bar{Q}}_2A_1A_2}^{w',\top}\right)\leq \Delta(V_{\user}^{w,\tilde{\bar{q}}_1,\tilde{\bar{q}}_2'},V_{\user}^{w',\tilde{\bar{q}}_1',\tilde{\bar{q}}_2'})=0,
\ee
where the inequality is due to the fact the state is diagonal in $\tilde{\bar{Q}}_1\tilde{\bar{Q}}_2A_1A_2$, and that $\tilde{\bar{Q}}_1\tilde{\bar{Q}}_2A_1A_2$ is part of the user's view for a setup with user query $\tilde{\bar{Q}}_1\tilde{\bar{Q}}_2$ and secure channels. By invoking the 0-database privacy of such a setup, there exist a $x'$ where the equality holds. We can therefore conclude that for any $x$, $r$ and keys ($s_1^{dec}$,$s_2^{enc}$,$s_3^{dec}$,$s_4^{enc}$), there exist an $x'$ such that

\be
\Delta(\xi_{\user\eve}^w,\xi_{\user\eve}^{w'})=0
\ee
Let us now examine the first term on the R.H.S. of Eq.~\eqref{dp_proof_eq1}. Likewise, we have that

\be
\Delta(\rho_{\user\eve}^w,\xi_{\user\eve}^w)\leq (1-p_{\fail})\Delta\left(\rho_{XR\bar{Q}_1\bar{Q}_2A_1A_2S_1S_2S_3S_4E}^{w,\top},\xi_{XR\bar{Q}_1\bar{Q}_2A_1A_2S_1S_2S_3S_4E}^{w,\top}\right).
\ee
We note that the following trace-preserving mappings are applied to $\bar{Q}_1\bar{Q}_2S_1S_2S_3S_4S_5S_6W$ to get $A_1A_2$,

\begin{multline*}
\bar{Q}_1S_1S_2 \overset{{\rm{partial\, trace}}}{\mapsto} \bar{Q}_1S_1^{\dec}S_2^\enc \overset{\rm{OTP}}{\mapsto} C_{\bar{Q}_1}S_1^{\dec}\overset{\rm{OTP}}{\mapsto}\tilde{\bar{Q}}_1, \quad
\tilde{\bar{Q}}_1WS_5 \overset{\baranswera}{\mapsto} A_1,\\
\bar{Q}_2S_3S_4 \overset{{\rm{partial\, trace}}}{\mapsto} \bar{Q}_2S_3^{\dec}S_4^\enc \overset{\rm{OTP}}{\mapsto} C_{\bar{Q}_2}S_3^{\dec}\overset{\rm{OTP}}{\mapsto}\tilde{\bar{Q}}_2, \quad
\tilde{\bar{Q}}_2WS_6 \overset{\baranswerb}{\mapsto} A_2,
\end{multline*}
Therefore, we obtain

\be
\Delta\left(\rho_{XR\bar{Q}_1\bar{Q}_2A_1A_2S_1S_2S_3S_4E}^{w,\top},\xi_{XR\bar{Q}_1\bar{Q}_2A_1A_2S_1S_2S_3S_4E}^{w,\top}\right)\leq\Delta\left(\rho_{XR\bar{Q}_1\bar{Q}_2S_1S_2S_3S_4S_5S_6WE}^{w,\top},\xi_{XR\bar{Q}_1\bar{Q}_2S_1S_2S_3S_4S_5S_6WE}^{w,\top}\right)
\ee
We note that $XR\bar{Q}_1\bar{Q}_2W$ is independent of $S_1S_2S_3S_4S_5S_6E$, and are common to both states. This thus gives us

\be
\Delta\left(\rho_{XR\bar{Q}_1\bar{Q}_2A_1A_2S_1S_2S_3S_4E}^{w,\top},\xi_{XR\bar{Q}_1\bar{Q}_2A_1A_2S_1S_2S_3S_4E}^{w,\top}\right)\leq\Delta\left(\rho_{S_1S_2S_3S_4S_5S_6E}^{\top},\xi_{S_1S_2S_3S_4S_5S_6E}^{\top}\right).
\ee
We can further partition $S_1S_2S_3S_4S_5S_6E$ into two parts, $S_5S_6$ and $S_1S_2S_3S_4E$, and note that $S_1S_2S_3S_4$ for both states are generated using real QKD protocol. With this, we may view the latter as some extended side-information $E'=S_1S_2S_3S_4E$. Then, using the security definition of QKD, we get that

\begin{multline}
\Delta(\rho_{\user\eve}^w,\xi_{\user\eve}^w)\leq (1-p_{\fail})\Delta\left(\rho_{S_1S_2S_3S_4S_5S_6E}^{\top},\xi_{S_1S_2S_3S_4S_5S_6E}^{\top}\right)\\
=(1-p_{\fail})\Delta(\rho_{S_5S_6E'}^{\top},\xi_{S_5S_6E'}^{\top})=\frac{1-p_{\fail}}{1-p_{\perp,12}}\Delta(\rhoQKD_{S_5S_6E'},\rhoideal_{S_5S_6E'})\leq \epscorr+\epssec,
\end{multline}
since $\xi^{\top}_{S_3S_4E'}$ is an ideal QKD output state conditioned on not aborting. Note that this is true for any $x'$. Combining the results, we conclude that there exist an $x'$ such that

\be
\Delta(\rho_{\user\eve}^w,\rho_{\user\eve}^{w'})\leq 2(\epscorr+\epssec).
\ee
\end{proof}

\begin{Theorem}
    A two-database one-round \textnormal{(0,0,0,0)}-secure classical SPIR protocol that uses $\varepsilon$-secure QKD keys in place of ideal keys, where $\varepsilon=\epscorr+\epssec$, is \textnormal{$4\varepsilon$}-protocol secret.
\end{Theorem}
\begin{proof}

We can define two intermediate states of $\eve$, each corresponding to a successive replacement of using real QKD keys with using an ideal QKD key. More specifically, $\xi_{\eve}^{x,w,1}$ is a setup replacing $(S_1,S_2)$ with ideal QKD keys, and $\xi_{\eve}^{x,w,2}$ corresponds to further replacing $(S_3,S_4)$ with ideal QKD keys. With these definitions, we can expand the trace distance in the protocol secrecy condition using the triangle inequality,

\begin{multline}
\label{ps_proof_eq1}
\Delta(\rho_{\eve}^{x,w},\rho_{\eve}^{x',w'})\leq \Delta(\rho_{\eve}^{x,w},\xi_{\eve}^{x,w,1})+\Delta(\xi_{\eve}^{x,w,1},\xi_{\eve}^{x,w,2})\\
+\Delta(\xi_{\eve}^{x,w,2},\xi_{\eve}^{x',w',2})+\Delta(\xi_{\eve}^{x',w',2},\xi_{\eve}^{x',w',1})+\Delta(\xi_{\eve}^{x',w',1},\rho_{\eve}^{x',w'}).
\end{multline}

We begin with examining the third term on the R.H.S. From Eq.~\eqref{eveview}, we get

\be
\Delta(\xi_{\eve}^{x,w,2},\xi_{\eve}^{x',w',2})=\Delta(\xi_{C_{Q_1}C_{Q_2}C_{A_1}C_{A_2}E}^{x,w,2},\xi_{C_{Q_1}C_{Q_2}C_{A_1}C_{A_2}E}^{x',w',2}).
\ee
Using the jointly convex property of trace distance, we obtain

\be
\Delta(\xi_{\eve}^{x,w,2},\xi_{\eve}^{x',w',2})\leq(1-p_{\fail})\Delta(\xi_{C_{Q_1}C_{Q_2}C_{A_1}C_{A_2}E}^{x,w,2,\top},\xi_{C_{Q_1}C_{Q_2}C_{A_1}C_{A_2}E}^{x',w',2,\top}).
\ee
Since ideal QKD keys are used between all parties, $C_{Q_1}$, $C_{Q_2}$, $C_{A_1}$, and $C_{A_2}$ are uniformly
distributed over $\vars{C}_{Q_1}$, $\vars{C}_{Q_2}$, $\vars{C}_{A_1}$, and $\vars{C}_{A_2}$ respectively conditioned on protocol not failing. With this, we can expand the trace distance to get

\begin{multline}
\Delta(\xi_{\eve}^{x,w,2},\xi_{\eve}^{x',w',2})\leq(1-p_{\fail})\\
\Delta\left(\sum_{c_{q_1}c_{q_2}c_{a_1}c_{a_2}}\frac{\Pi_{C_{Q_1}C_{Q_2}C_{A_1}C_{A_2}}(c_{q_1}c_{q_2}c_{a_1}c_{a_2})}{\abs{\vars{C}_{Q_1}}\abs{\vars{C}_{Q_2}}\abs{\vars{C}_{A_1}}\abs{\vars{C}_{A_2}}}\otimes\sigma_E,\right.\\ \left.\sum_{c_{q_1}c_{q_2}c_{a_1}c_{a_2}}\frac{\Pi_{C_{Q_1}C_{Q_2}C_{A_1}C_{A_2}}(c_{q_1}c_{q_2}c_{a_1}c_{a_2})}{\abs{\vars{C}_{Q_1}}\abs{\vars{C}_{Q_2}}\abs{\vars{C}_{A_1}}\abs{\vars{C}_{A_2}}}\otimes\sigma_E\right)=0.
\end{multline}

Let us now examine the second term on the R.H.S. of Eq.~\eqref{ps_proof_eq1}. We first obtain

\be
\Delta(\xi_{\eve}^{x,w,1},\xi_{\eve}^{x,w,2})\leq(1-p_{\fail})\Delta(\xi_{C_{Q_1}C_{Q_2}C_{A_1}C_{A_2}E}^{x,w,1,\top},\xi_{C_{Q_1}C_{Q_2}C_{A_1}C_{A_2}E}^{x,w,2,\top})
\ee
Since ideal QKD keys $(S_1,S_2)$ are used, $C_{Q_1}$ and $C_{A_1}$ are uniformly distributed over $\vars{C}_{Q_1}$ and $\vars{C}_{A_1}$ respectively, conditioned on the protocol not failing. With this, we can expand the trace distance to get

\begin{multline}
\Delta(\xi_{\eve}^{x,w,1,\top},\xi_{\eve}^{x,w,2,\top})\leq(1-p_{\fail})\Delta\left(\xi_{C_{Q_2}C_{A_2}E}^{x,w,1,\top}\otimes\sum_{c_{q_1}c_{a_1}}\frac{\Pi_{C_{Q_1}C_{A_1}}(c_{q_1}c_{a_1})}{\abs{\vars{C}_{Q_1}}\abs{\vars{C}_{A_1}}},\right.\\
\left.\xi_{C_{Q_2}C_{A_2}E}^{x,w,2,\top}\otimes\sum_{c_{q_1}c_{a_1}}\frac{\Pi_{C_{Q_1}C_{A_1}}(c_{q_1}c_{a_1})}{\abs{\vars{C}_{Q_1}}\abs{\vars{C}_{A_1}}}\right).
\end{multline}
We note that the following trace preserving map can be applied to $Q_2WS_3S_4S_6$ to obtain $C_{Q_2}C_{A_2}$,

\[
Q_2S_3S_4 \overset{{\rm{partial\, trace}}}{\mapsto} Q_2S_3^{\dec}S_4^\enc \overset{\rm{OTP}}{\mapsto} C_{Q_2}S_3^{\dec}\overset{\rm{OTP}}{\mapsto}\tilde{Q}_2, \quad
\tilde{Q}_2WS_6 \overset{\baranswerb}{\mapsto} A_2, \quad A_2S_3 \overset{\textrm{partial\,trace}}{\mapsto} A_2S_3^{\enc}\overset{\rm{OTP}}{\mapsto} C_{A_2}.
\]
Therefore, we get

\be
\Delta(\xi_{\eve}^{x,w,1,\top},\xi_{\eve}^{x,w,2,\top})\leq(1-p_{\fail})\Delta(\xi_{Q_2WS_3S_4S_6E}^{x,w,1,\top},\xi_{Q_2WS_3S_4S_6E}^{x,w,2,\top}).
\ee
We first note that $Q_2W$ is independent of $S_3S_4S_6E$, and is common to both terms, thus resulting in

\be
\Delta(\xi_{Q_2WS_3S_4S_6E}^{x,w,1,\top},\xi_{Q_2WS_3S_4S_6E}^{x,w,2,\top})\leq\Delta(\xi_{S_3S_4S_6E}^{1,\top},\xi_{S_3S_4S_6E}^{2,\top}).
\ee
We can further partition $S_3S_4S_6E$ into two parts, $S_3S_4$ and $S_6E$, and note that $S_6$ for both states are generated using real QKD protocol. With this, we may view the latter as some extended side-information $E'=S_6E$. Then, using the security definition of QKD, we get that

\begin{multline}
\Delta(\xi_{\eve}^{x,w,1,\top},\xi_{\eve}^{x,w,2,\top})\leq(1-p_{\fail})\Delta(\xi_{S_3S_4S_6E}^{1,\top},\xi_{S_3S_4S_6E}^{2,\top})\\
\leq \frac{1-p_{\fail}}{1-p_{\abort,U2}}\Delta(\rhoQKD_{S_3S_4E'},\rhoideal_{S_3S_4E'})\leq \epscorr+\epssec.
\end{multline}

We next examine the first term on the R.H.S. of Eq.~\eqref{ps_proof_eq1}. We first obtain 

\be
\Delta(\rho_{\eve}^{x,w},\xi_{\eve}^{x,w,1})\leq(1-p_{\fail})\Delta(\sigma_{C_{Q_1}C_{Q_2}C_{A_1}C_{A_2}E}^{x,w,\top},\xi_{C_{Q_1}C_{Q_2}C_{A_1}C_{A_2}E}^{x,w,1,\top})
\ee
We note that the following map can be applied on $Q_1Q_2WS_1S_2S_3S_4S_5S_6$ to obtain $C_{Q_1}C_{Q_2}C_{A_1}C_{A_2}$,

\begin{multline*}
Q_1S_1S_2 \overset{{\rm{partial\, trace}}}{\mapsto} Q_1S_1^{\dec}S_2^\enc \overset{\rm{OTP}}{\mapsto} C_{Q_1}S_1^{\dec}\overset{\rm{OTP}}{\mapsto}\tilde{Q}_1, \quad
\tilde{Q}_1WS_5 \overset{\baranswera}{\mapsto} A_1, \quad A_1S_1 \overset{\rm{partial\,trace}}{\mapsto} A_1S_1^{\enc}\overset{\rm{OTP}}{\mapsto} C_{A_1},\\
Q_2S_3S_4 \overset{{\rm{partial\, trace}}}{\mapsto} Q_2S_3^{\dec}S_4^\enc \overset{\rm{OTP}}{\mapsto} C_{Q_2}S_3^{\dec}\overset{\rm{OTP}}{\mapsto}\tilde{Q}_2, \quad
\tilde{Q}_2WS_6 \overset{\baranswerb}{\mapsto} A_2, \quad A_2S_3 \overset{\rm{partial\,trace}}{\mapsto} A_2S_3^{\enc}\overset{\rm{OTP}}{\mapsto} C_{A_2}.
\end{multline*}
Therefore, we get

\be
\Delta(\sigma_{C_{Q_1}C_{Q_2}C_{A_1}C_{A_2}E}^{x,w,\top},\xi_{C_{Q_1}C_{Q_2}C_{A_1}C_{A_2}E}^{x,w,1,\top})\leq\Delta(\sigma_{Q_1Q_2WS_1S_2S_3S_4S_5S_6E}^{x,w,\top},\xi_{Q_1Q_2WS_1S_2S_3S_4S_5S_6E}^{x,w,1,\top}).
\ee
Since $Q_1Q_2W$ is independent on $S_1S_2S_3S_4S_5S_6E$, and is common to both terms (with same $x$ and $w$), we obtain

\be
\Delta(\sigma_{Q_1Q_2WS_1S_2S_3S_4S_5S_6E}^{x,w,\top},\xi_{Q_1Q_2WS_1S_2S_3S_4S_5S_6E}^{x,w,1,\top})\leq\Delta(\sigma_{S_1S_2S_3S_4S_5S_6E}^{\top},\xi_{S_1S_2S_3S_4S_5S_6E}^{1,\top}).
\ee
We can further partition $S_1S_2S_3S_4S_5S_6E$ into two parts, $S_1S_2$ and $S_3S_4S_5S_6E$, and note that $S_3S_4S_5S_6$ is common for both states. With this, we may view the latter as some extended side-information $E'=S_3S_4S_5S_6E$. Then, we get that

\be
\Delta(\rho_{\eve}^{x,w},\xi_{\eve}^{x,w,1})\leq(1-p_{\fail})\Delta(\sigma_{S_1S_2E'}^{\top},\xi_{S_1S_2E'}^{1,\top})\leq\frac{1-p_{\fail}}{1-p_{\perp,U1}}\Delta(\rhoQKD_{S_1S_2E'},\rhoideal_{S_1S_2E'})\leq\epscorr+\epssec.
\ee

Combining the results, we obtain

\be
\Delta(\rho_{\eve}^{x,w},\rho_{\eve}^{x',w'})\leq 4\varepsilon.
\ee
\end{proof}

\section{$\btwoprot$ Protocol}
\label{B2ProtApp}

For simplicity, we consider a database with size $n=m^3$, with one-bit database entries, $W=(w_1,\dots,w_n)\in\{0,1\}^n$. We label the entries with index $X=(X^1,X^2,X^3)$, where $X^i\in\{1,\dots,m\}$, for $i=1,2,3$. The user has a source of local randomness labelled by $R=(R_{s},R_d)$. $R_{s}$ consists of three random subsets, $R_{s}^i\subseteq\{1,\dots,m\}$ (which can be expressed as a random $m$-bit vector as well), and $R_d$ is a set of three values, $R_d^i\in\{1,\dots,m\}$. Furthermore, we label the pre-shared keys, between the two data centres, $K_3K_4$, by $(U,T,Y,Z)$, which are used for CDS. We also define the notation

\begin{equation*}
    S\triangle\{j\}=\begin{cases}
        S\setminus\{j\} & j\in S\\
        S\cup\{j\} & j\notin S
    \end{cases}
\end{equation*}
for a set $S$.

We first define the query used in the $\btwoprot$ protocol. The user first selects a desired index $x=(x^1,x^2,x^3)$, and generates the local random values $R_{s}$ and $R_d$. Query to data centre 1 is simply $Q_1=(Q_{1,s},Q_{1,d})$, where $Q_{1,s}=R_{s}$ and $Q_{1,d}=R_d$. For the query to data centre 2, the user has to compute $Q_{2,d}^i\equiv x^i-R_d^i (\text{mod}\,m)$, and $Q_{2,s}=R_{s}^i\triangle\{x^i\}$. The query is thus $Q_2=(Q_{2,s},Q_{2,d})$. Essentially, the user encodes his desired index in both the set query as the only element that is contained exclusively in $Q_{1,s}$ or $Q_{2,s}$ and the index query as the sum of $Q_{1,d}$ and $Q_{2,d}$ modulo $m$. 

The data centre answers consist of 8 portions, which are labelled by index $\sigma=\{0,1\}^3$, and one portion responsible for CDS to ensure that the user provides valid queries. The keys used for masking the responses are $U$ and $T$. $U$ consists of 3 random bits, $U^i$, $T$ consists of 8 bits, $T^{\sigma}$, of which 7 are random, and the final bit is chosen to ensure $\bigoplus_{\sigma}T^{\sigma}=0$. Keys that are used for CDS are $Y$ and $Z$. $Y$ is a set of 6 vectors of length $m$, $Y^{\sigma}$, for $\sigma=\{001,010,100,011,101,110\}$, and $Z$ is a set of 3 vectors of length $m$, $Z^i$. Data centre 1 then computes the answers for $j\in\{1,\dots,m\}$ and $i=1,2,3$,

\begin{equation*}
    \begin{split}
        A^{000}=&\left[\bigoplus_{k\in(Q_{1,s}^1,Q_{1,s}^2,Q_{1,s}^3)}w_k\right]\oplus T^{000}\\
        A^{100}_j=&\left[\bigoplus_{k\in(Q_{1,s}^1\triangle\{j\},Q_{1,s}^2,Q_{1,s}^3)}w_k\right]\oplus Y^{100}_{j-Q_{1,d}^{1}}\oplus T^{100}\\
        A^{010}_j=&\left[\bigoplus_{k\in(Q_{1,s}^1,Q_{1,s}^2\triangle\{j\},Q_{1,s}^3)}w_k\right]\oplus Y^{010}_{j-Q_{1,d}^{2}}\oplus T^{010}\\
        A^{001}_j=&\left[\bigoplus_{k\in(Q_{1,s}^1,Q_{1,s}^2,Q_{1,s}^3\triangle\{j\})}w_k\right]\oplus Y^{001}_{j-Q_{1,d}^{3}}\oplus T^{001}\\
        A^{\CDS}_i=&\left[\bigoplus_jI^{Q_{1,s}^i}_jZ^i_j\right]\oplus U^i,
    \end{split}
\end{equation*}
where $I^S$ is the indicator function of set $S$ (i.e. $I^S_j=1$ if $j\in S$ and $I^S_j=0$ if $j\notin S$). The computed values, together with three additional bits $Y^{011}_{Q_{1,d}^1}$, $Y^{101}_{Q_{1,d}^2}$, and $Y^{110}_{Q_{1,d}^3}$, forms the answer $A_1$. Data centre 2 computes the answer

\begin{equation*}
        \begin{split}
        A^{111}=&\left[\bigoplus_{k\in(Q_{2,s}^1,Q_{2,s}^2,Q_{2,s}^3)}w_k\right]\oplus T^{111}\\
        A^{011}_j=&\left[\bigoplus_{k\in(Q_{2,s}^1\triangle\{j\},Q_{2,s}^2,Q_{2,s}^3)}w_k\right]\oplus Y^{011}_{j-Q_{2,d}^{1}}\oplus T^{011}\oplus Z^1_j\\
        A^{101}_j=&\left[\bigoplus_{k\in(Q_{2,s}^1,Q_{2,s}^2\triangle\{j\},Q_{2,s}^3)}w_k\right]\oplus Y^{101}_{j-Q_{2,d}^{2}}\oplus T^{101}\oplus Z^2_j\\
        A^{110}_j=&\left[\bigoplus_{k\in(Q_{2,s}^1,Q_{2,s}^2,Q_{2,s}^3\triangle\{j\})}w_k\right]\oplus Y^{110}_{j-Q_{2,d}^{3}}\oplus T^{110}\oplus Z^3_j\\
        A^{\CDS'}_i=&\left[\bigoplus_jI^{Q_{2,s}^i}_jZ^i_j\right]\oplus U^i.
    \end{split}
\end{equation*}
The above values, together with three extra bits, $Y^{100}_{Q_{2,d}^1}$, $Y^{010}_{Q_{2,d}^2}$, and $Y^{001}_{Q_{2,d}^3}$, forms the answer $A_2$.

The decoding function is obtained by simply performing an XOR on some of the answer bits received by the user. If the user is honest, the correct value of $\hat{w}_x$ can be obtained from the decoding function. Firstly, by taking the sum of the CDS answers, we can retrieve the value of $Z_x$ using

\begin{equation*}
    Z^i_{x^i}=A_i^{\CDS}\oplus A_i^{\CDS'}.
\end{equation*}
Since $Q_{1,d}^i+Q_{2,d}^i\equiv x^i (\text{mod}\,m)$, the dependency of $A^{\sigma}$ on $Y^{\sigma}$ can be removed by choosing $j=x^i$ for the appropriate $i$. The final decoding would thus be

\begin{multline*}
\hat{w}_x=\left[(A^{100}_{x^1}\oplus Y^{100}_{Q_{2,d}^1})\oplus(A^{011}_{x^1}\oplus Y^{011}_{Q_{1,d}^1}) \oplus Z^1_{x^1}\right]\oplus\left[(A^{010}_{x^2}\oplus Y^{010}_{Q_{2,d}^2})\oplus(A^{101}_{x^2}\oplus Y^{101}_{Q_{1,d}^2})\right]\\
\oplus\left[(A^{001}_{x^3}\oplus Y^{001}_{Q_{2,d}^3})\oplus(A^{110}_{x^3}\oplus Y^{110}_{Q_{1,d}^3})\right]\oplus A^{111}\oplus A^{000}.
\end{multline*}


\reftitle{References}

\externalbibliography{yes}
\bibliography{QKDSPIRManu}

\end{document}